\documentclass[]{spie}  

\usepackage[]{graphicx,floatrow,sidecap}
\usepackage{wrapfig,placeins,afterpage}

\newfloatcommand{capbtabbox}{table}[][\FBwidth]

\newcommand{\mka}{Mn-K$_\alpha$ }

\usepackage{supertabular}

\usepackage[dvips]{color}

\title{Microcalorimeter pulse analysis by means of principle component decomposition}

\author{
	C.P.~de~Vries\supit{a},
        R.M.~Schouten\supit{a},
        J.~van~der~Kuur\supit{a},
        L.~Gottardi\supit{a} and
        H.~Akamatsu\supit{a}
\skiplinehalf
\supit{a} SRON Netherlands Institute for Space Research, 
          Sorbonnelaan~2, 3584~CA~Utrecht, Netherlands \\
       }

\begin{document}                                                               
\maketitle 

\begin{abstract}
The X-ray integral field unit for the Athena mission consists of a microcalorimeter
transition edge sensor pixel array. Incoming photons generate pulses which
are analyzed in terms of energy, in order to assemble the X-ray spectrum.
Usually this is done by means of optimal filtering in either time or frequency domain.  

In this paper we investigate an alternative method by means of principle component analysis.
This method attempts to find the main components of an orthogonal set of functions to describe the data.

We show, based on simulations, what the influence of various instrumental effects is on this type of 
analysis. We compare analyses both in time and frequency domain. Finally we apply these analyses on
real data, obtained via frequency domain multiplexing readout.
 
\end{abstract}

\section{INTRODUCTION}
The proposed X-ray integral field unit (X-IFU)\cite{xifu2013} for the Athena\cite{Athena2013} mission uses a
microcalorimeter array as a 2-dimensional detection device, capable of retrieving high resolution spectra
for each imaging element. Each pixel contains a transition edge sensor (TES) which records the
energy pulses of incoming X-ray photons. The XIFU detector will have of order 4000 pixels in the array.
At SRON we develop a TES-pixel readout scheme by means of frequency domain multiplexing \cite{Hartog2011}.

\begin{SCfigure}
\includegraphics[width=0.40\textwidth,clip]{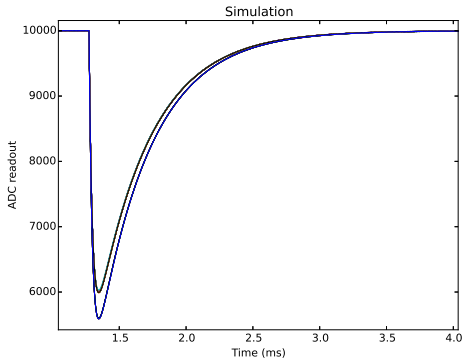}
{\caption{\small Shape of ideal model pulses, without any systematic instrumental effects, but with
some independent noise added. The energy of the pulses is
drawn from the expected \mka distribution}
\label{fig:ideal_pulse}}
\end{SCfigure}

The recorded pulses will have to be analyzed in terms of incoming photon energy. Usually this is done by means
of optimal filtering\cite{szymkowiak1993}, fitting a pulse shape in the time, or frequency
domain using appropriate weights based on the noise spectrum.
In this paper we use an alternate way of processing by means of principle component analysis (PCA). PCA attempts
to find the main components in an orthogonal set of shapes (eigenvectors) which describe the measured pulses. 
The main eigenvalues, or projections of the pulse onto the main eigenvectors, will have a relation with the 
incoming photon energy. Ideally only one main component should represent energy and the other components
should represent other effects present in the data, but in practice several parameters are mixed in the different components. 
Projecting a data vector onto an eigenvector (the in-product) can also be seen as multiplying the data
vector with a weights vector. As such the PCA method can be seen as a modification of the optimal filtering
method, in the sense that PCA derives a best set of weights instead of taking them directly from noise spectrum,
for a (set of) component(s) which measure the energy. 

Previously Busch et al.\cite{busch2015} have reported on the application of
this method to real microcalorimeter data pulses of 
photons, in time domain DC readout mode. They found this method to be quite promising to obtain the best 
energy resolution.

Here we report on the application of this method to simulated data, in order to investigate the relation of 
different effects present in the data on the components found by PCA. In addition we apply PCA to microcalorimeter
data obtained at SRON from a multi-pixel array by means of frequency domain multiplexing (FDM).
We compare analyses both in time and frequency domain. We aim to minimize the number of relevant components,
ideally to obtain the smallest set of components which relate to the energy of the photons, in
other words, to maximize the isolation of a component representing the energy.

\section{DATA SIMULATION}

\begin{figure}[b]
\centering
\begin{floatrow}
\ffigbox{
\includegraphics[width=0.95\hsize,clip]{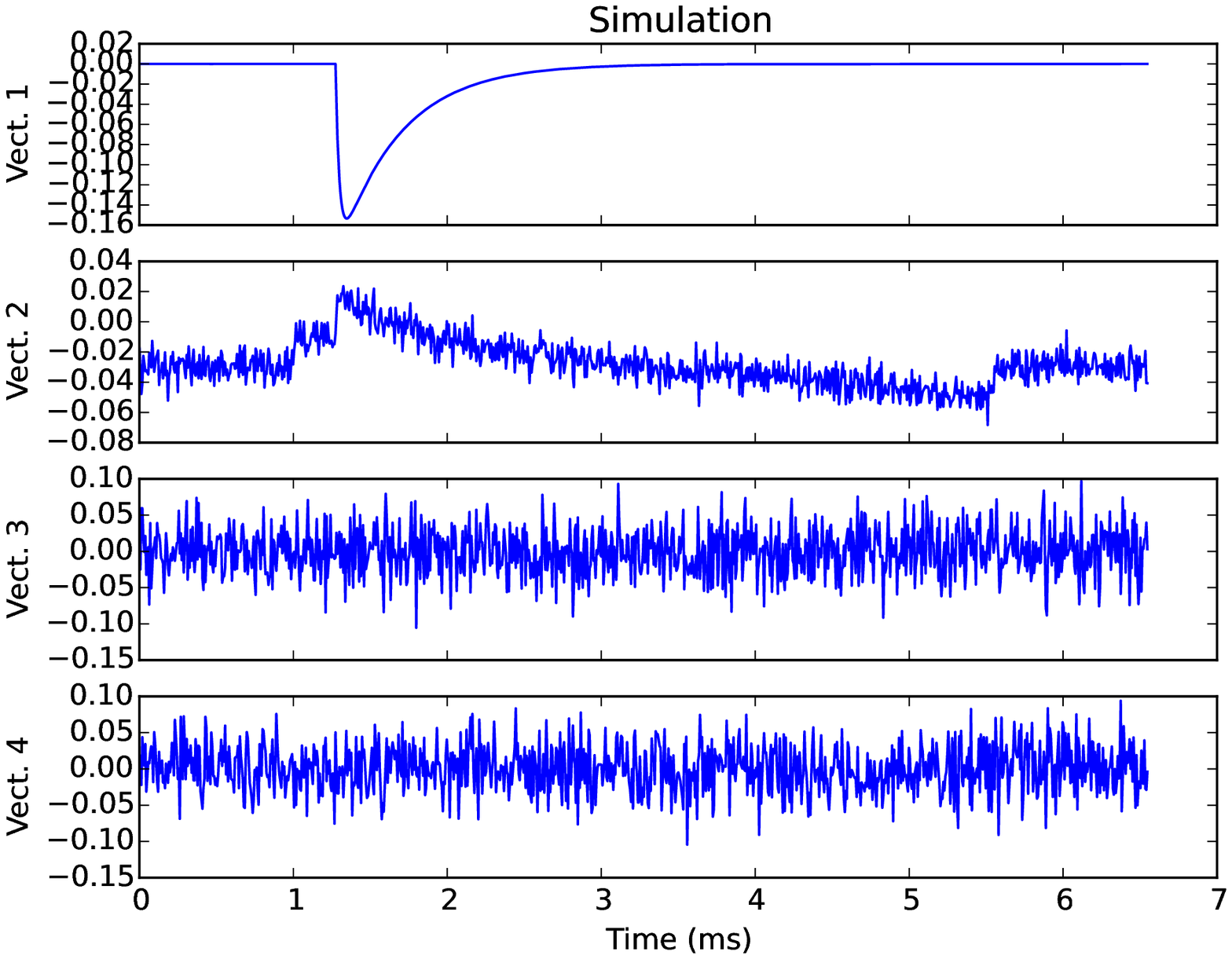}
}{
\caption{\small PCA components of ideal pulses. Only the first component, which
resembles the pulse shape, and which corresponds to energy is relevant.}
\label{fig:ideal_components}
} 
\ffigbox{
\includegraphics[width=0.90\hsize,clip]{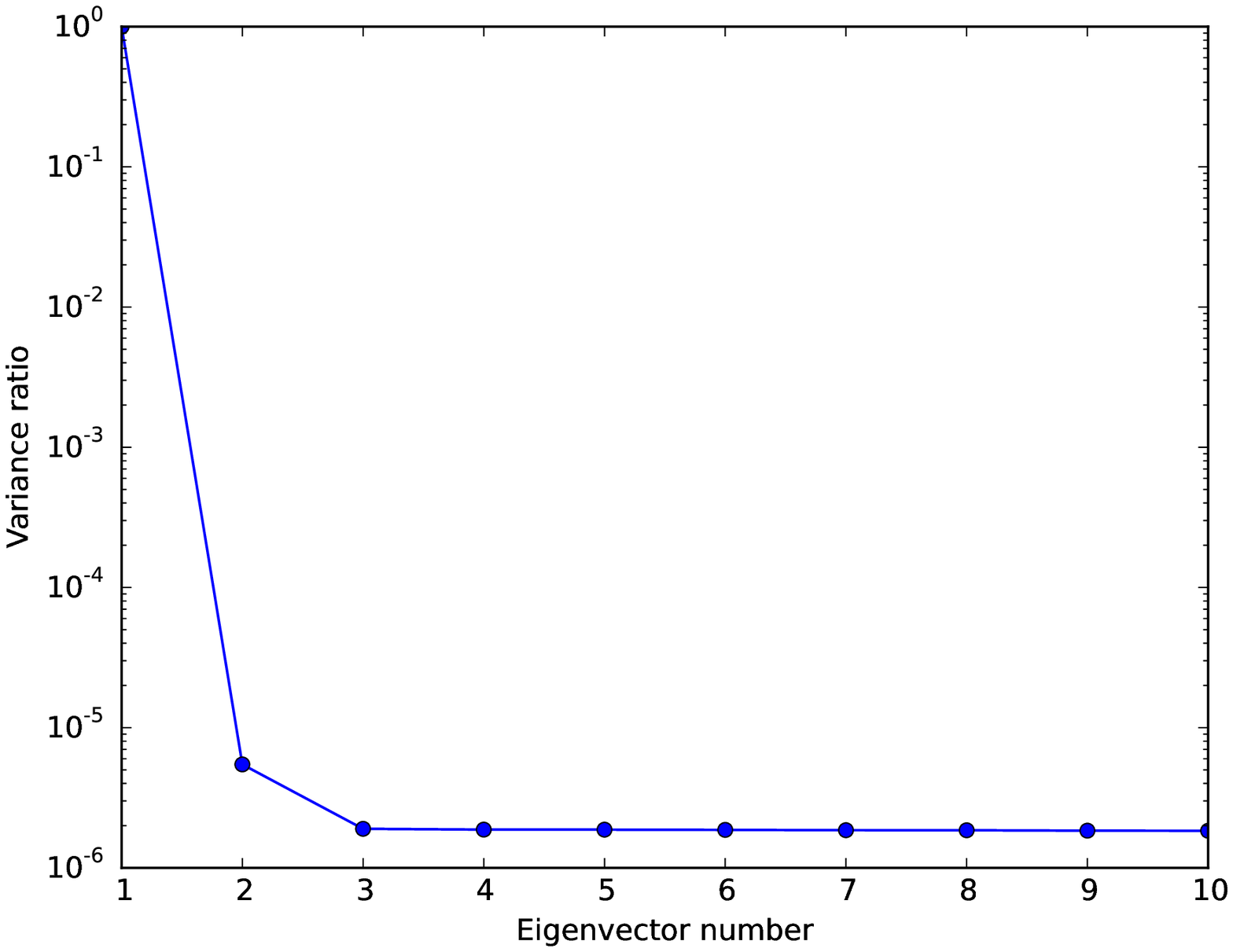}
}{
\caption{\small Variance ratio of the ideal pulse components, which corresponds to the relative
importance. Only the first component matters.}
\label{fig:ideal_varratio}
} 
\end{floatrow}
\end{figure}

For the simulations, energies are drawn at random from a distribution function for the \mka 
X-ray line as published by Holzer\cite{holzer1997}. Pulses are generated using exponential rise and fall times
representing the SRON detectors and are scaled according to energy. This signal represents the ideal case. On top
of this, instrumental effects are added:
\begin{itemize}
\item{The baseline level is modified using a drifting linear slope and offset}
\item{A drifting gain factor which changes the pulse level plus a partial offset}
\item{A changing pulse start time, correlated with the energy, to mimic the
      constant trigger threshold with respect to the changing amplitude of the pulse}
\item{A non-linear, saturation relation between input and 'recorded' signal}
\item{Added noise to the signal}
\end{itemize} 
These modifications can be given any magnitude to study the effect of these disturbances on the components found
by the PCA. The results of the PCA are analyzed in terms of the fitted instrumental resolution (convolution
width) on top of the Holzer\cite{holzer1997} natural line width. The fitting method uses the
C-stat\cite{wheaton1995,cash1979}
statistics to cope with the limited number of counts in each spectral bin.

\begin{figure}[tb]
\centering
\begin{floatrow}
\ffigbox{
\includegraphics[width=0.90\hsize,clip]{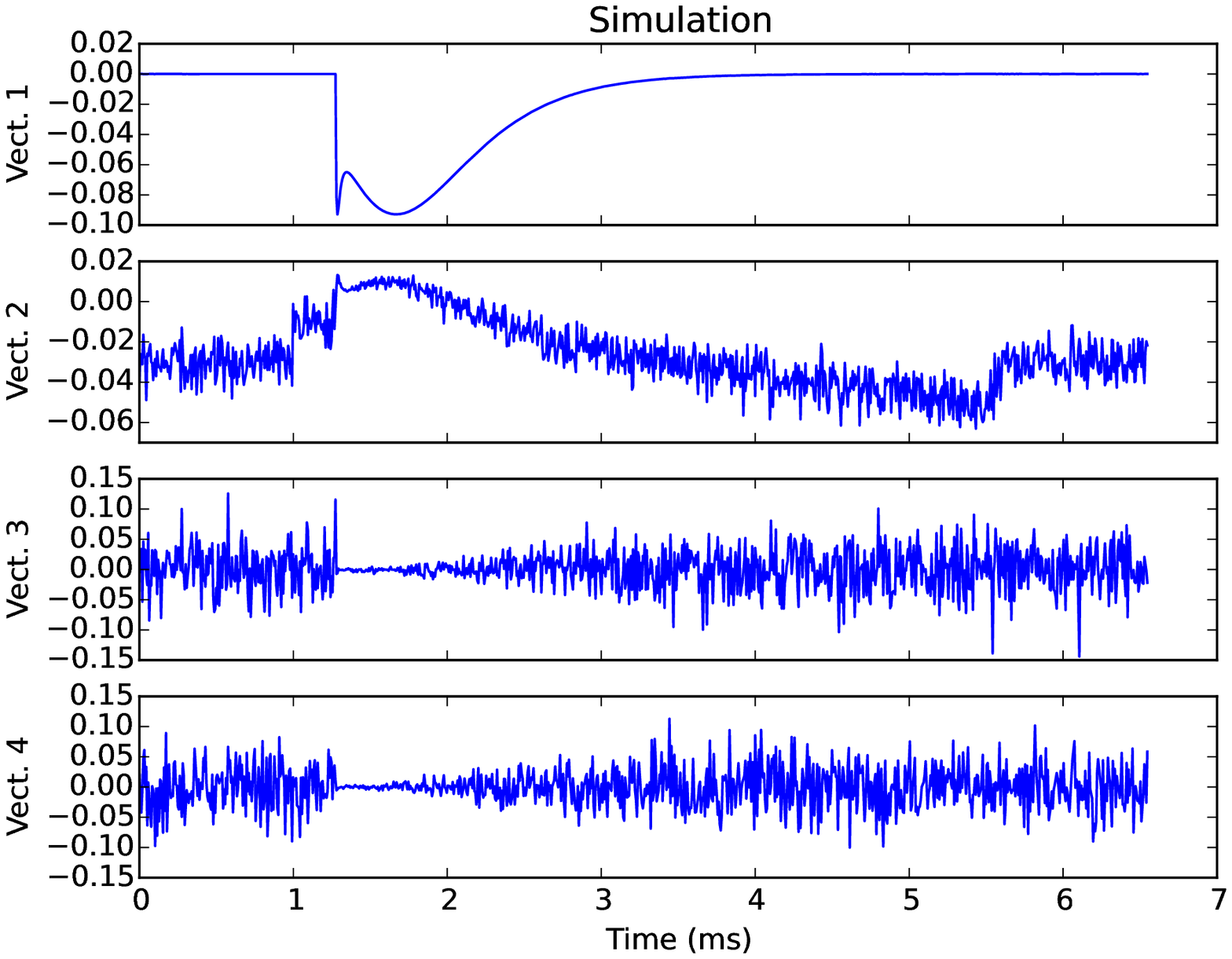}
}{
\caption{\small PCA components of saturated pulses.}
\label{fig:sat2000_components}
} 
\ffigbox{
\includegraphics[width=0.90\hsize,clip]{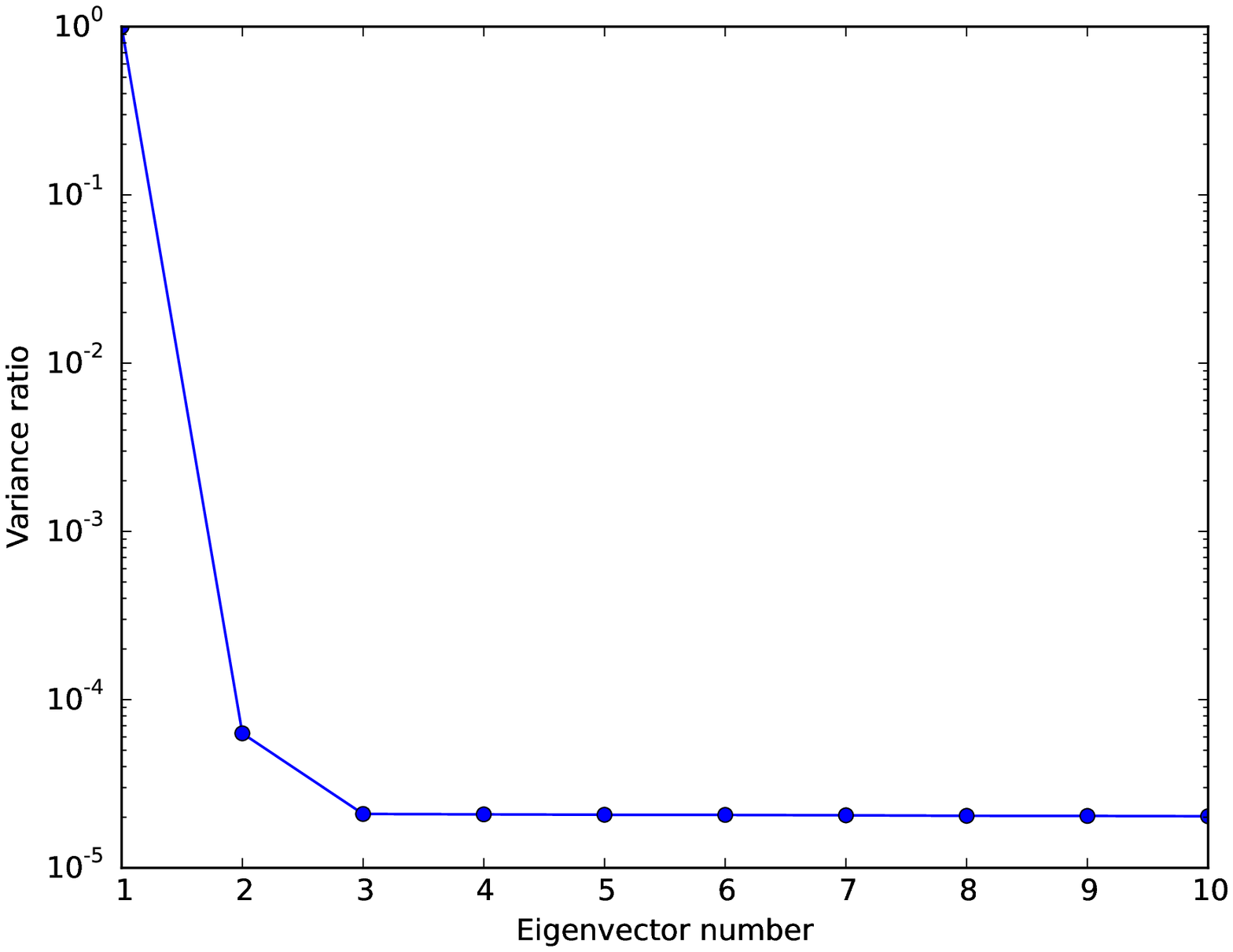}
}{
\caption{\small Variance ratio of the saturated pulse components}
\label{fig:sat2000_varratio}
} 
\end{floatrow}

\begin{floatrow}
\ffigbox{
\includegraphics[width=0.90\hsize,clip]{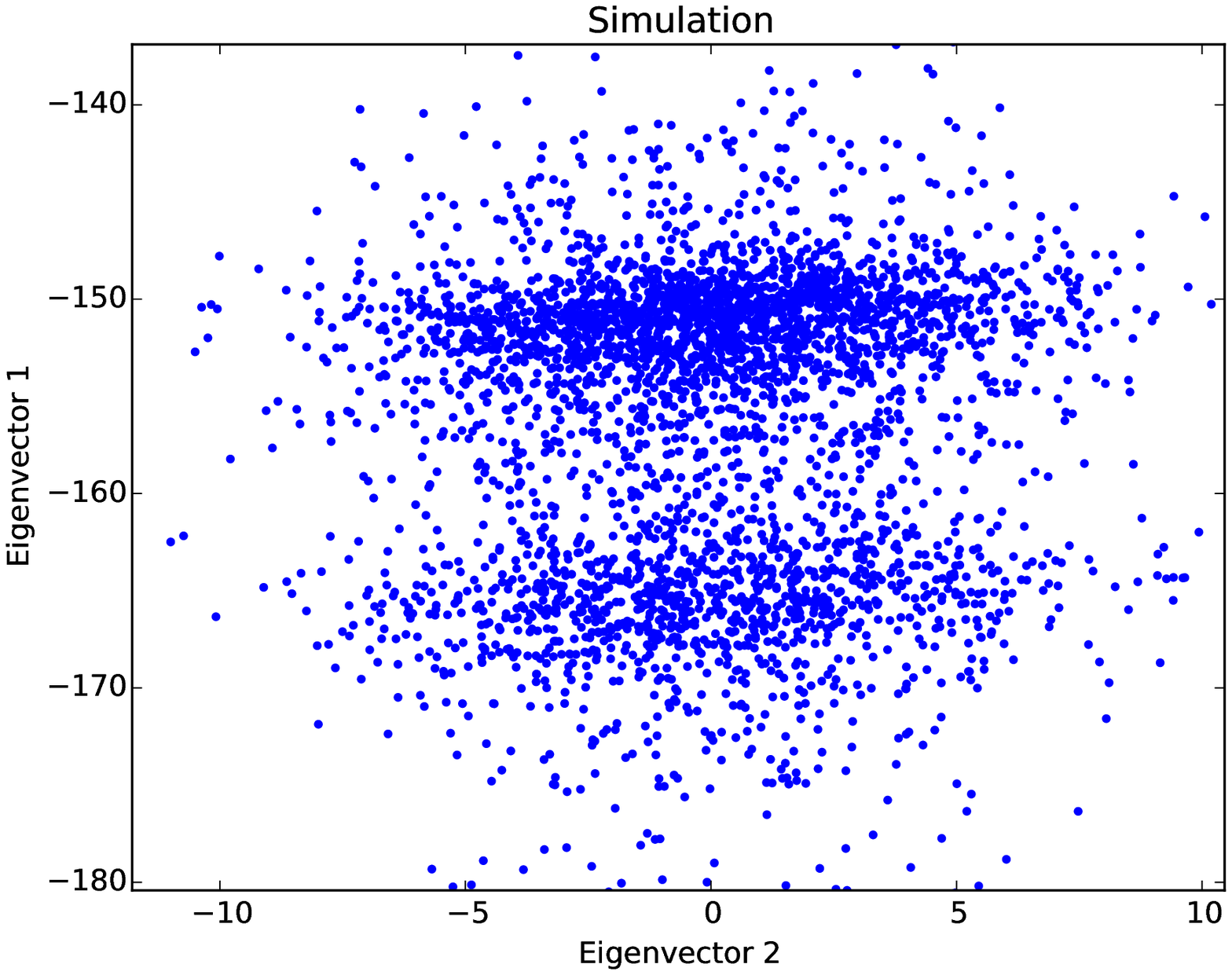}
}{
\caption{\small Correlation between the first two PCA components.}
\label{fig:sat2000_v1v2}
} 
\ffigbox{
\includegraphics[width=0.90\hsize,clip]{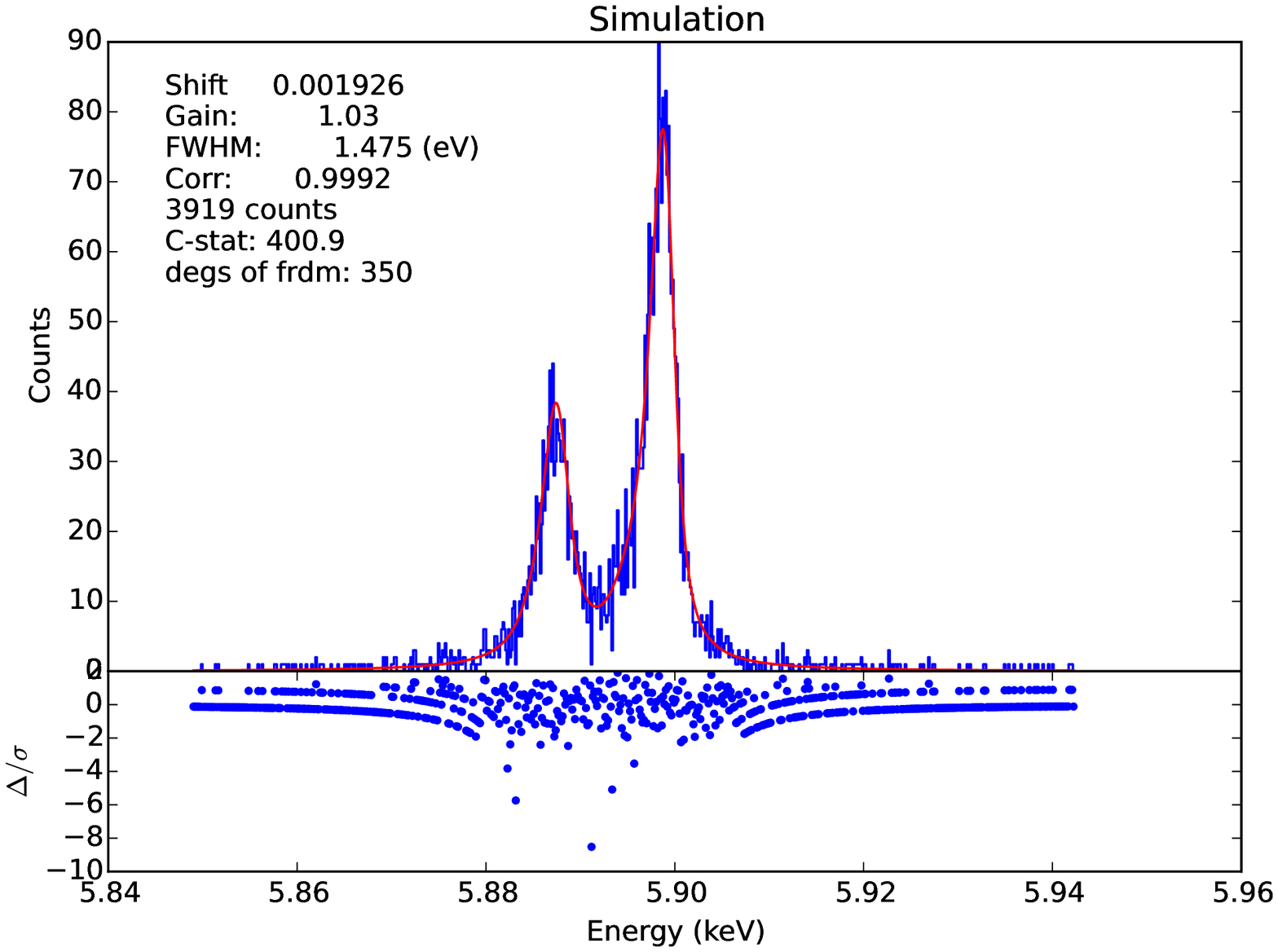}
}{
\caption{\small Fit of the instrumental resolution on the \mka line.}
\label{fig:sat2000_lfit}
} 
\end{floatrow}

\end{figure}

Figure~\ref{fig:ideal_pulse} shows the ideal modelled pulse shape. PCA analysis 
(fig~\ref{fig:ideal_components}) shows that only one component is relevant (fig~\ref{fig:ideal_varratio}).
This first component appears like a copy of the basic pulse shape.
There appears to be some structure in the second component, but its relevance is only $5\times10^{-6}$ with
respect to the first component. In addition, this relevance can be lowered to any arbitrary low number by
decreasing the noise level.
This is to be expected, since the only parameters which discriminates the pulses is the energy, and no further
instrumental effects are present. The fitted instrumental resolution on the \mka lines is consistent with
no instrumental broadening effect (0.0~eV).  

When saturating effects (fig~\ref{fig:sat2000_components}) in the signal are introduced,
things start to look different. Relevance of the
second component now rises to $6\times10^{-5}$ (fig.~\ref{fig:sat2000_varratio}).
The shape of the first component also changes, and there
is a slight correlation between first and second components (fig.~\ref{fig:sat2000_v1v2}). Instrumental
resolution is now fitted at 1.5~eV.

\begin{figure}[tb]

\begin{floatrow}
\ffigbox{
\includegraphics[width=0.90\hsize,clip]{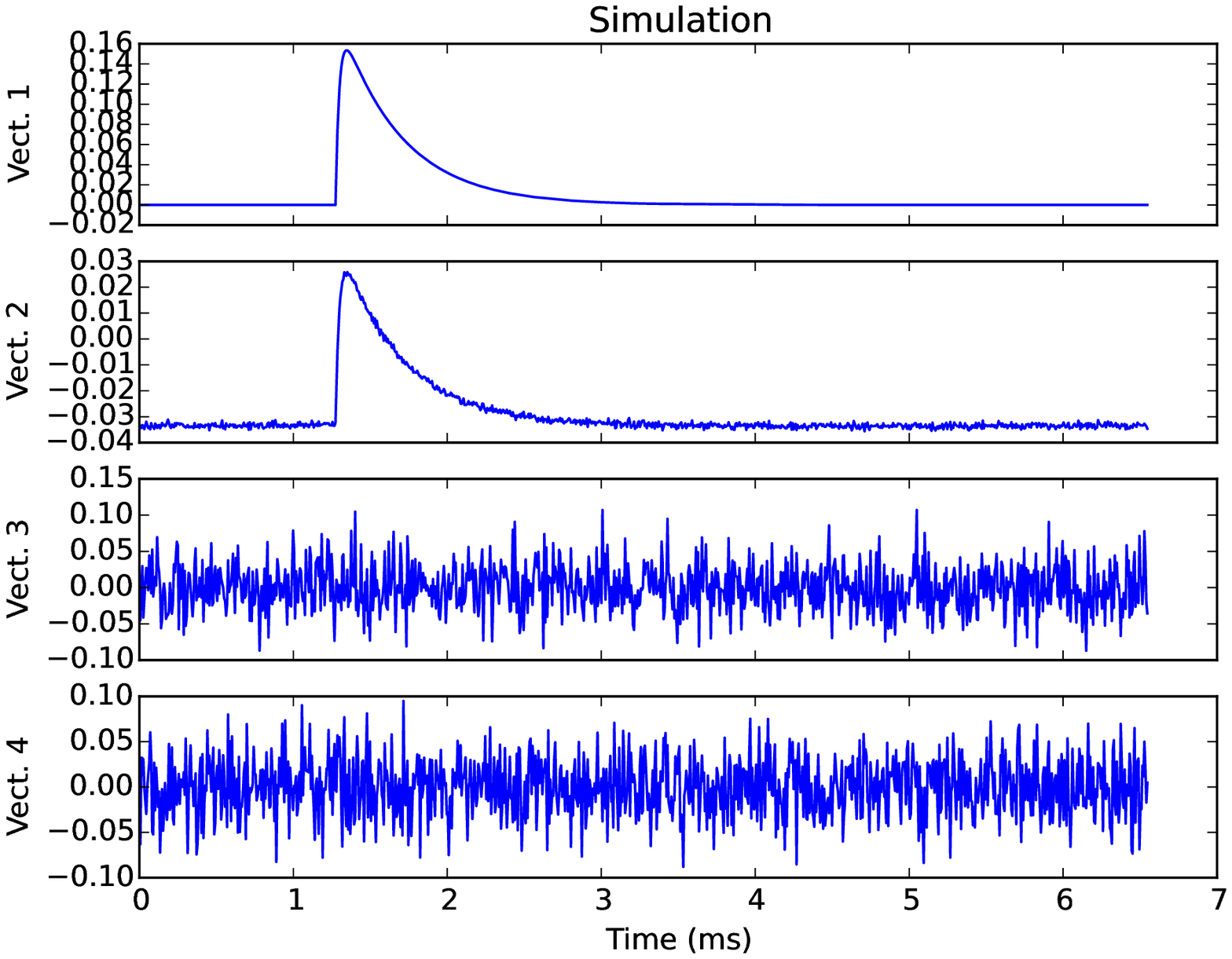}
}{
\caption{\small PCA components of pulses subject to gain drifts.}
\label{fig:gsdrift6_components}
} 
\ffigbox{
\includegraphics[width=0.90\hsize,clip]{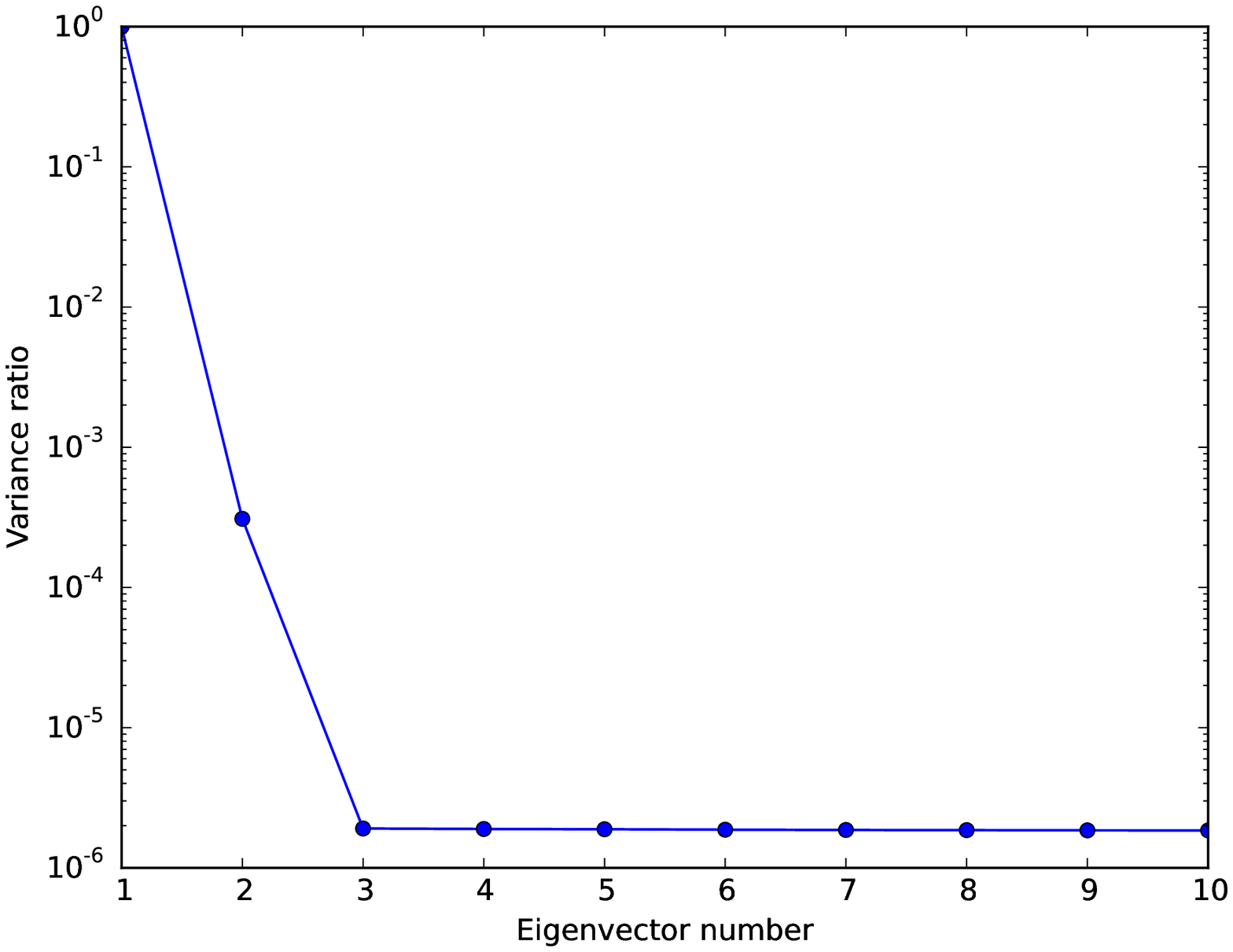}
}{
\caption{\small Variance ratio of the gain drift pulses}
\label{fig:gsdrift6_varratio}
} 
\end{floatrow}

\begin{floatrow}
\ffigbox{
\includegraphics[width=0.90\hsize,clip]{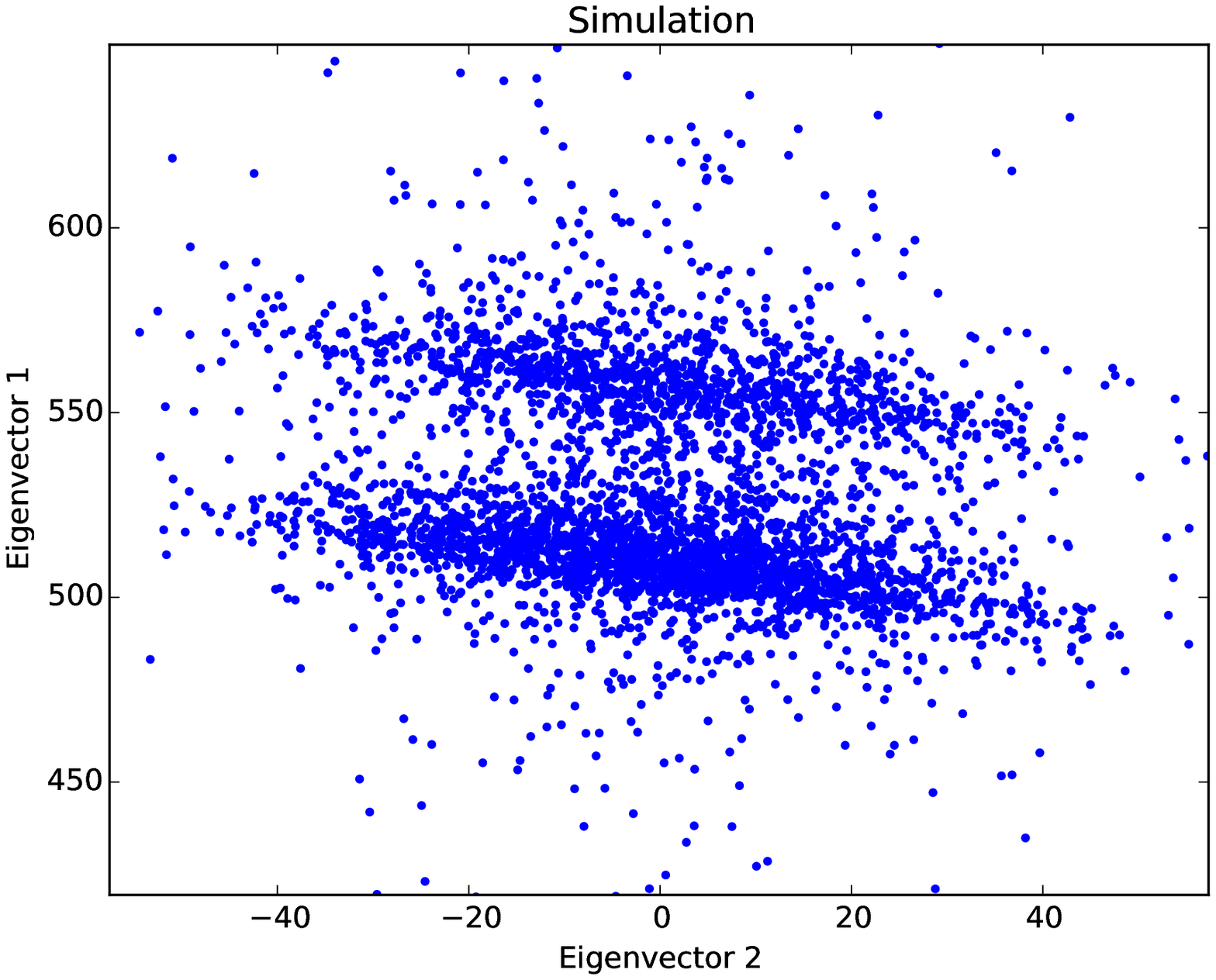}
}{
\caption{\small Correlation between the first two PCA components.}
\label{fig:gsdrift6_v1v2}
} 
\ffigbox{
\includegraphics[width=0.90\hsize,clip]{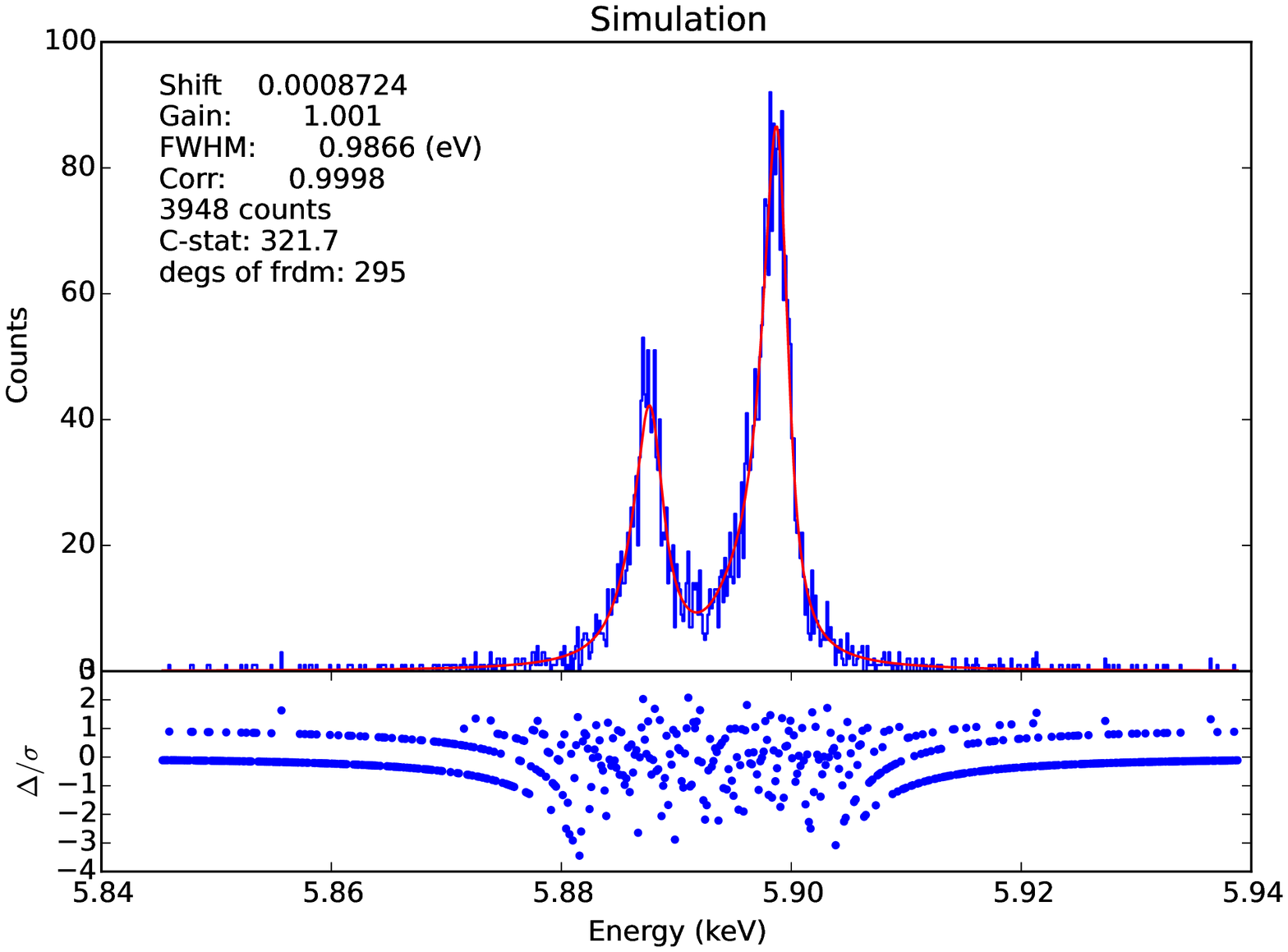}
}{
\caption{\small Fit of the instrumental resolution on the \mka line.}
\label{fig:gsdrift6_lfit}
} 
\end{floatrow}
\end{figure} 
\clearpage

Gain drifts show their effects in figures~\ref{fig:gsdrift6_components} to \ref{fig:gsdrift6_lfit}.
The second component also matters here at a level of $3\times10^{-4}$, but the PCA is quite able to
cope with this effect. Final instrumental resolution is fitted at about 1~eV.

\begin{SCfigure}
\includegraphics[width=0.40\textwidth,clip]{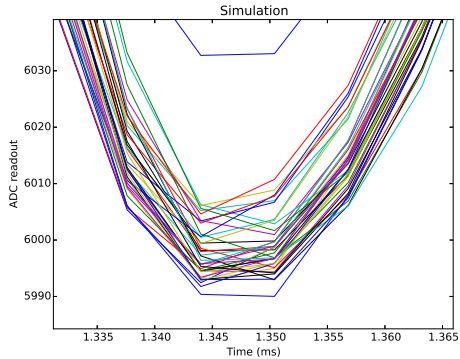}
{\caption{\small Differences between identical pulses, as shown on the top of the pulse, due to the
unequal moments of sampling the pulses. 
}
\label{fig:jitt_sample}}
\end{SCfigure} 

\begin{figure}[bt]
\centering
\begin{floatrow}
\ffigbox{
\includegraphics[width=0.90\hsize,clip]{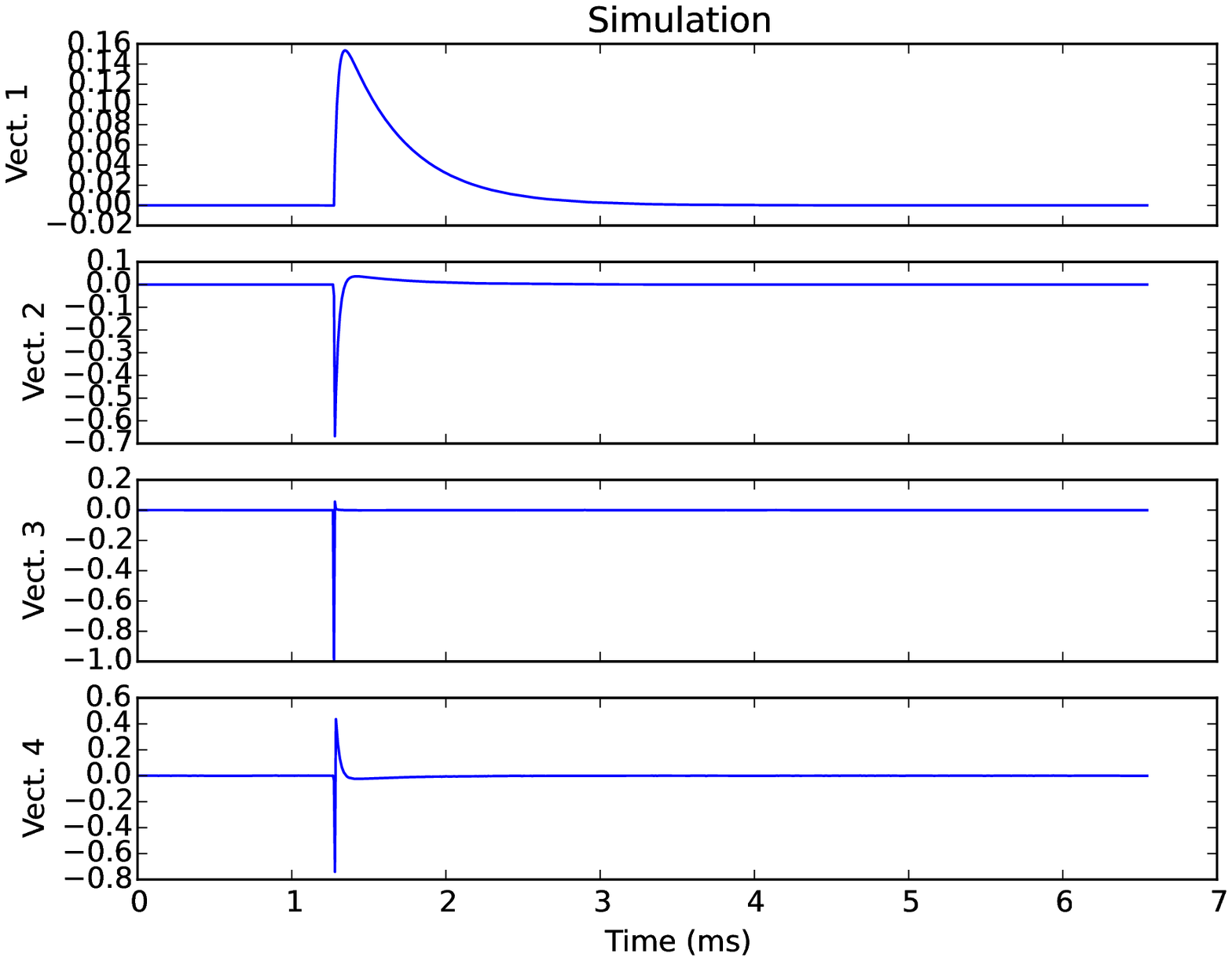}
}{
\caption{\small PCA components of pulses subject to sampling jitter.}
\label{fig:jitt_components}
} 
\ffigbox{
\includegraphics[width=0.90\hsize,clip]{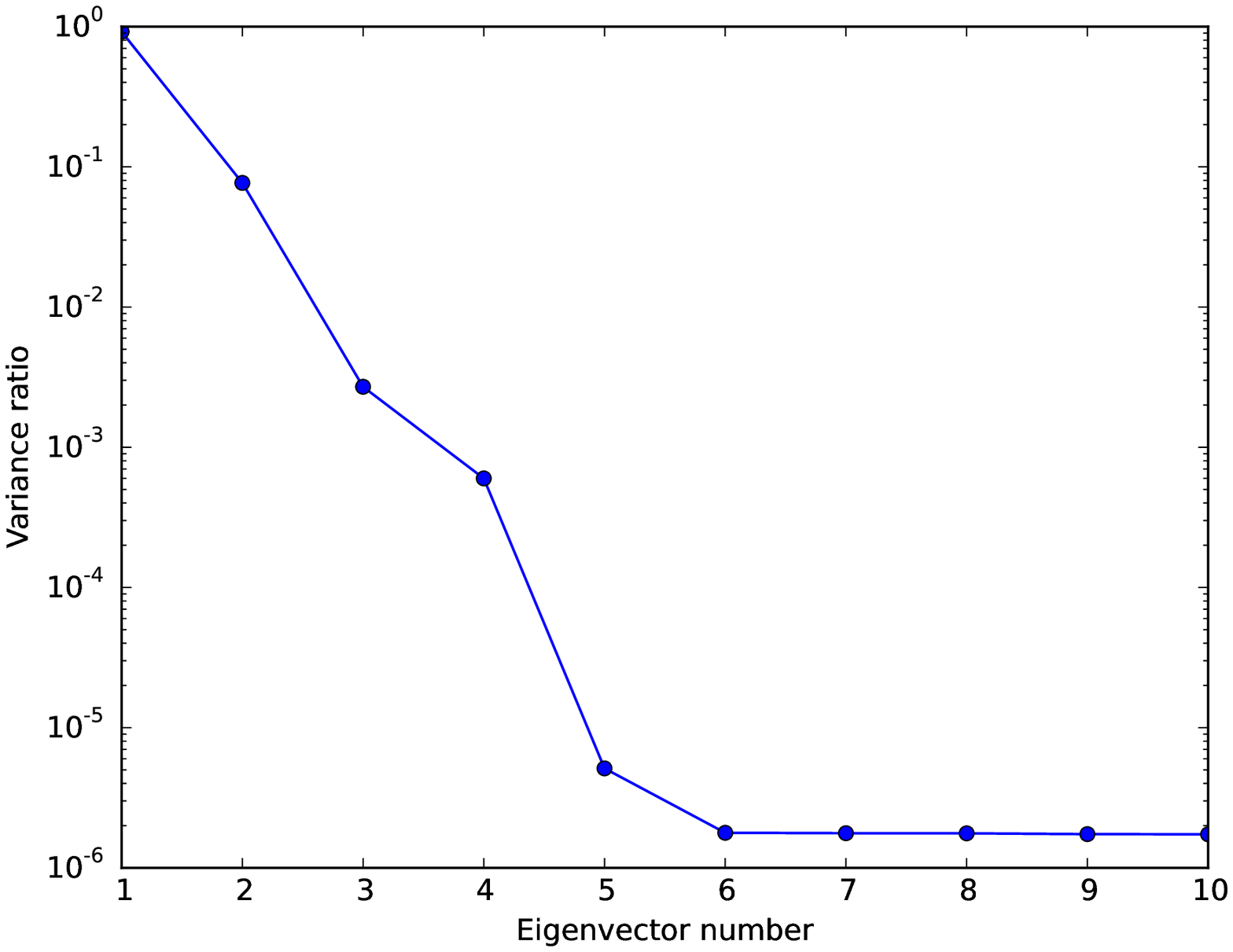}
}{
\caption{\small Variance ratio of the sampling jitter pulses}
\label{fig:jitt_varratio}
} 
\end{floatrow}

\begin{floatrow}
\ffigbox{
\includegraphics[width=0.90\hsize,clip]{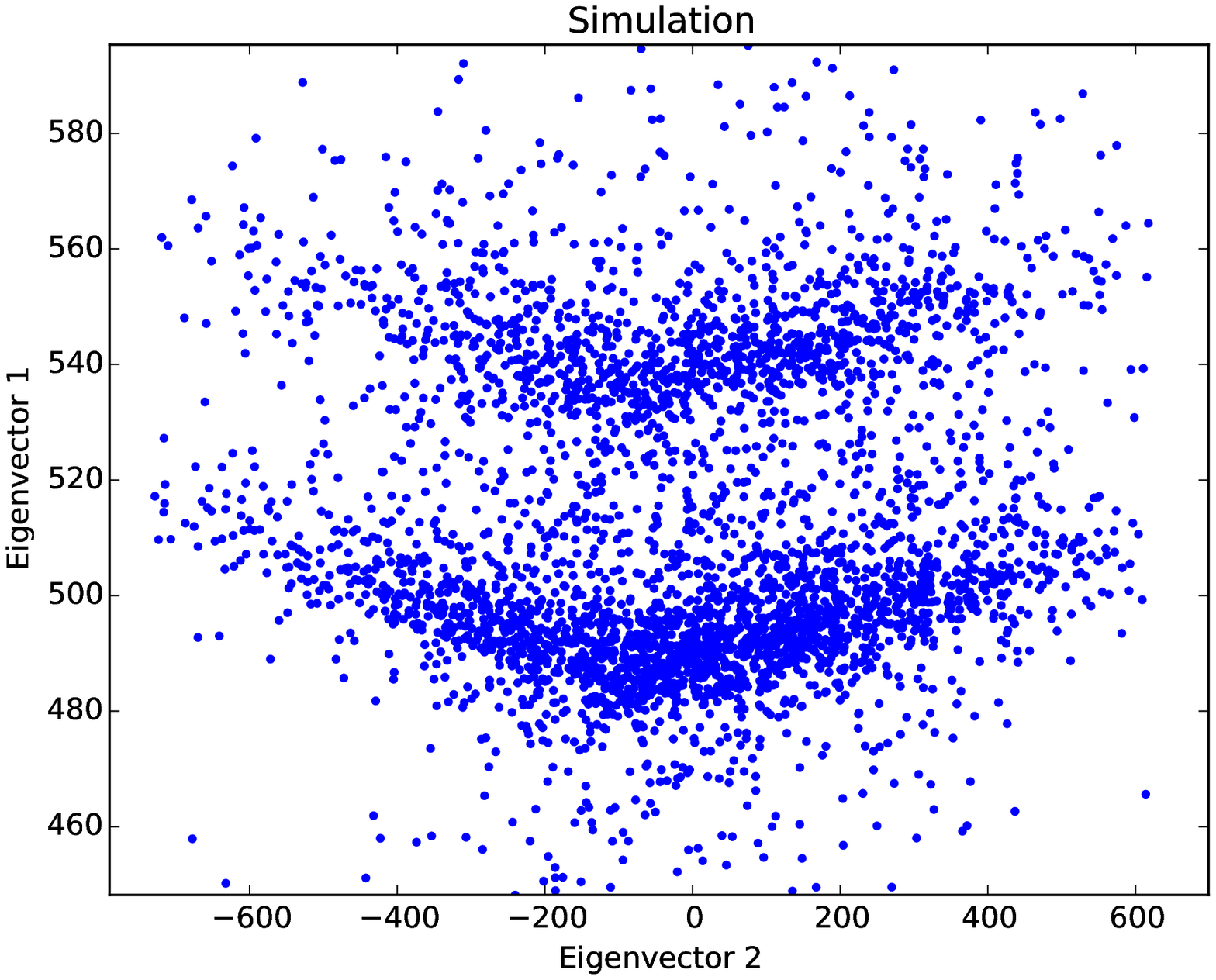}
}{
\caption{\small Correlation between the first two PCA components.}
\label{fig:jitt_v1v2}
} 
\ffigbox{
\includegraphics[width=0.90\hsize,clip]{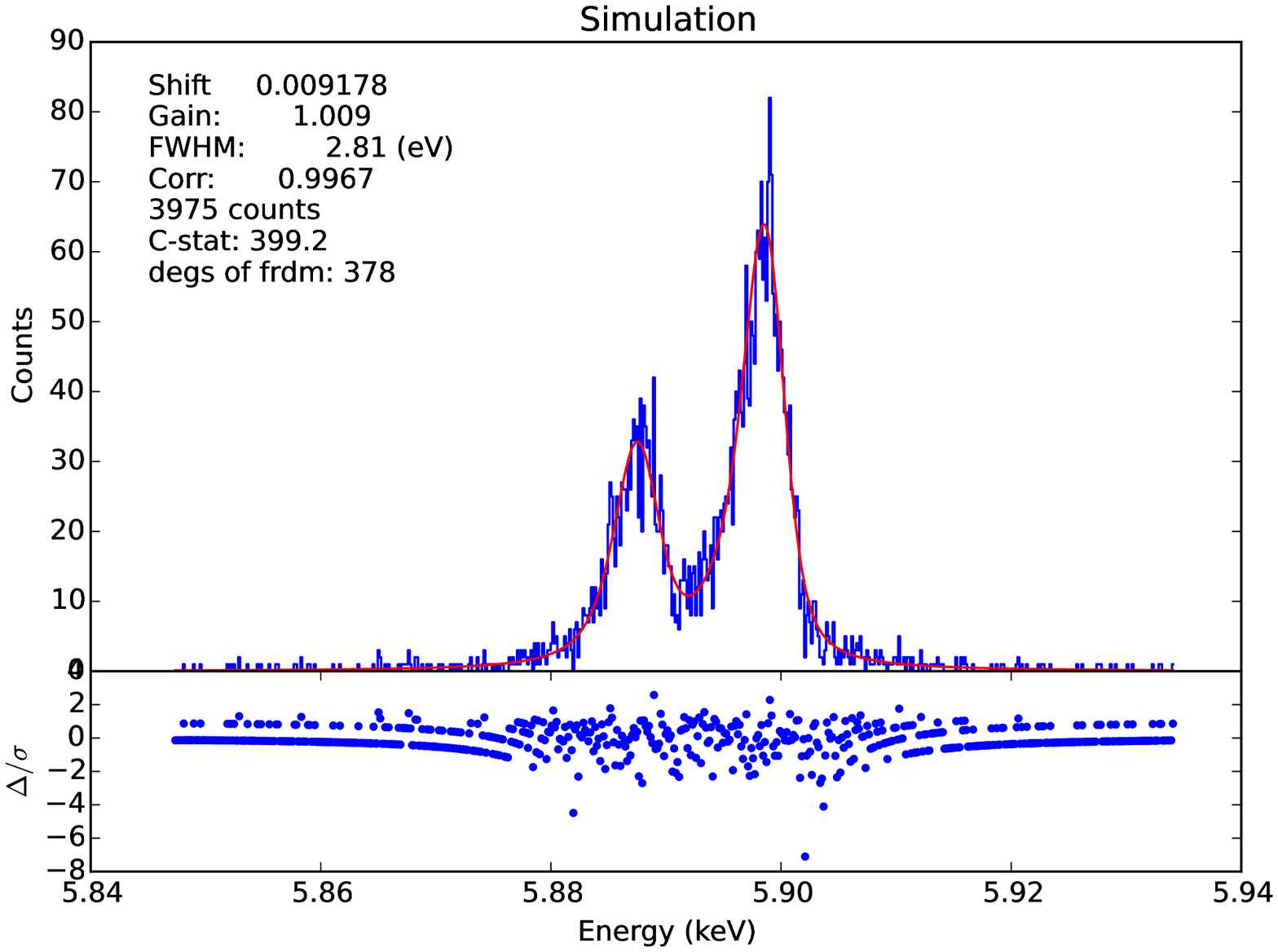}
}{
\caption{\small Fit of the instrumental resolution on the \mka line.}
\label{fig:jitt_lfit}
} 
\end{floatrow}
\end{figure}

Final effect is a drift of the pulse with respect to the sampling times. When the pulse is not
highly over sampled, the actual pulse shape seems to differ (see figure\ref{fig:jitt_sample}). This 
does not mean that the pulse is under sampled, since proper interpolation can restore the
correct pulse form at any position. However this requires significant additional computing resources.
With this effect the direct PCA in time domain breaks down. Figures \ref{fig:jitt_components} to
\ref{fig:jitt_lfit} show the effects for very limited
drifts. More than two components now become relevant. There is also clear correlation between the first
and second components and this correlation is not linear (fig.~\ref{fig:jitt_v1v2}). Instrumental resolution
suffers (2.8 eV, fig.~\ref{fig:jitt_lfit}).

\afterpage{\clearpage}
 
\section{FREQUENCY DOMAIN ANALYSIS}

\begin{figure}[tb]
\centering
\begin{floatrow}
\ffigbox{
\includegraphics[width=0.90\hsize,clip]{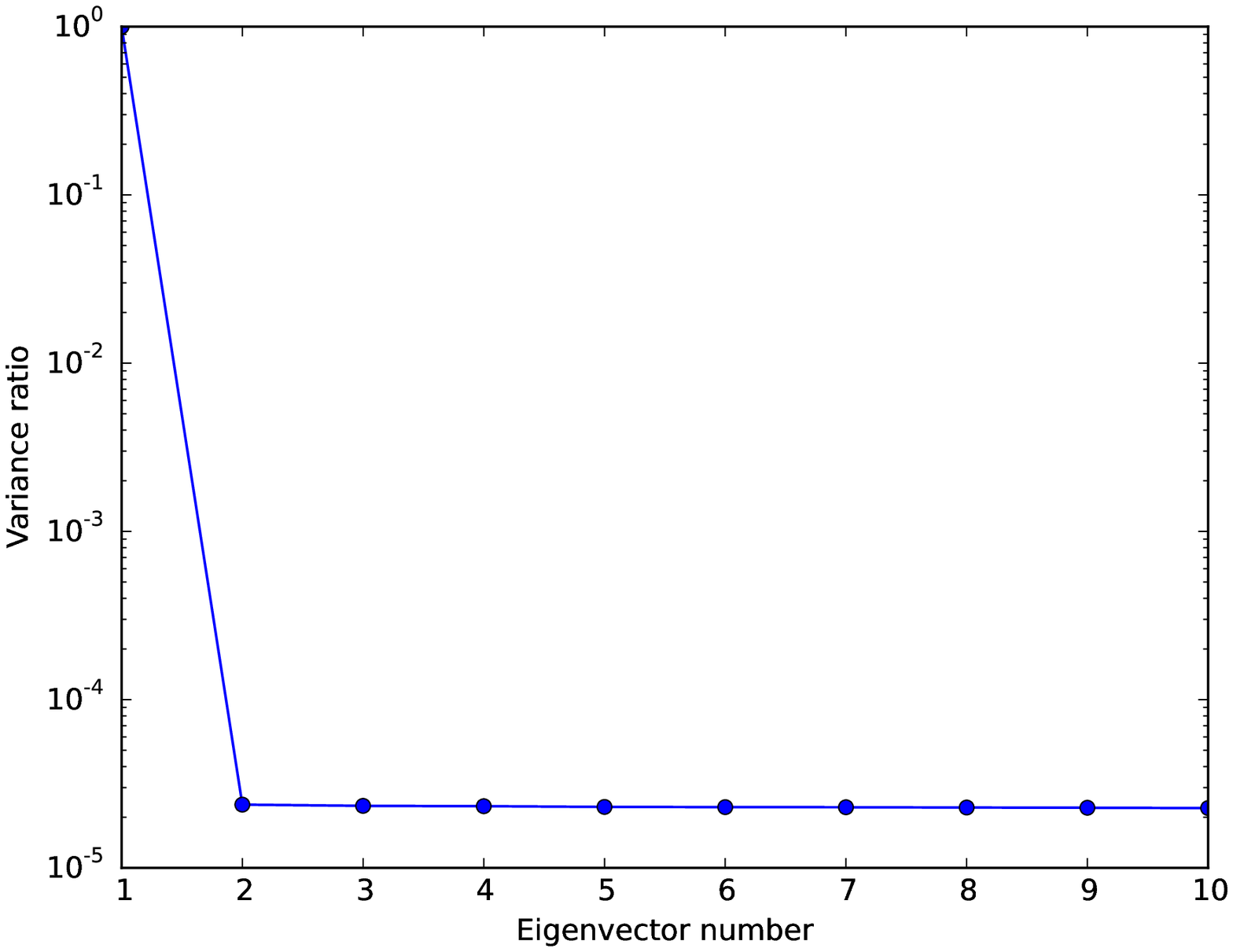}
}{
\caption{\small Relative importance of components for the saturated pulses,
analyzed in the frequency domain. Compare figure~\ref{fig:sat2000_varratio} for the time domain.
The second component is now significantly lower and at the same level as the higher components.}
\label{fig:f_sat2000_varratio}
} 
\ffigbox{
\includegraphics[width=0.90\hsize,clip]{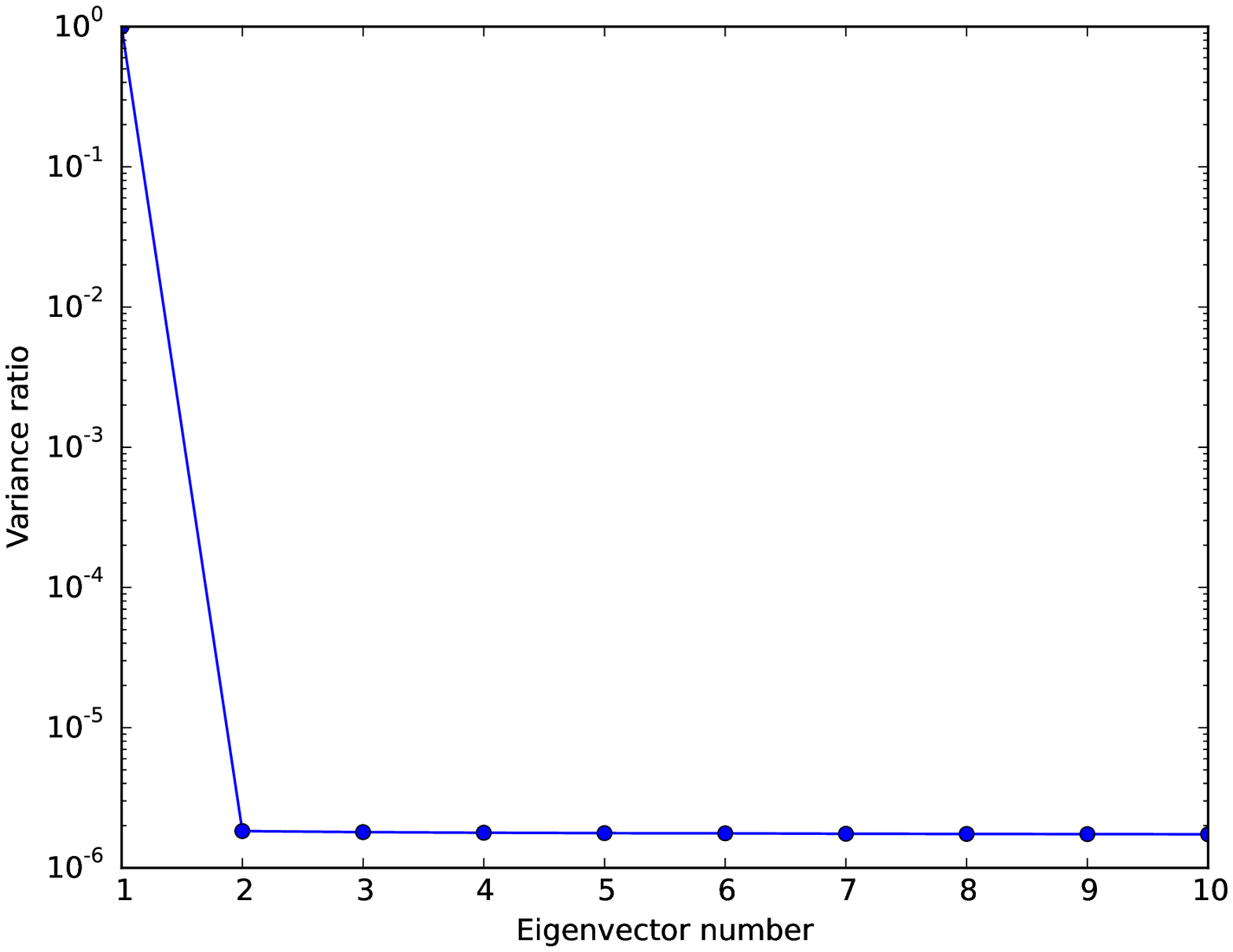}
}{
\caption{\small Relative importance of components for the gain drifted pulses,
analyzed in the frequency domain. Compare figure~\ref{fig:gsdrift6_varratio} for the time domain.
The second component is now significantly lower and at the same level as the higher components.}
\label{fig:f_gsdrift6_varratio}
} 
\end{floatrow}
\end{figure}

Apart from analysis in the time domain, PCA can also be done in the frequency domain.
Using the power spectrum of the pulse instead of the pulse itself, has the advantage that
the phase of the pulse becomes irrelevant. This is confirmed in the simulation analysis.
Using the same simulation parameters as before, in the case of saturation and gain drifts
the PCA analyses in the frequency domain do not make much difference. Although in the
relative importance of the various components only the first component plays a role and
other components are all at the same low level (around or below $1\times10^{-5}$, see
figures~\ref{fig:f_sat2000_varratio} and \ref{fig:f_gsdrift6_varratio}) the 
final instrumental resolutions obtained are similar to the time domain case above. 

\begin{figure}[bt]
\centering
\begin{floatrow}
\ffigbox{
\includegraphics[width=0.90\hsize,clip]{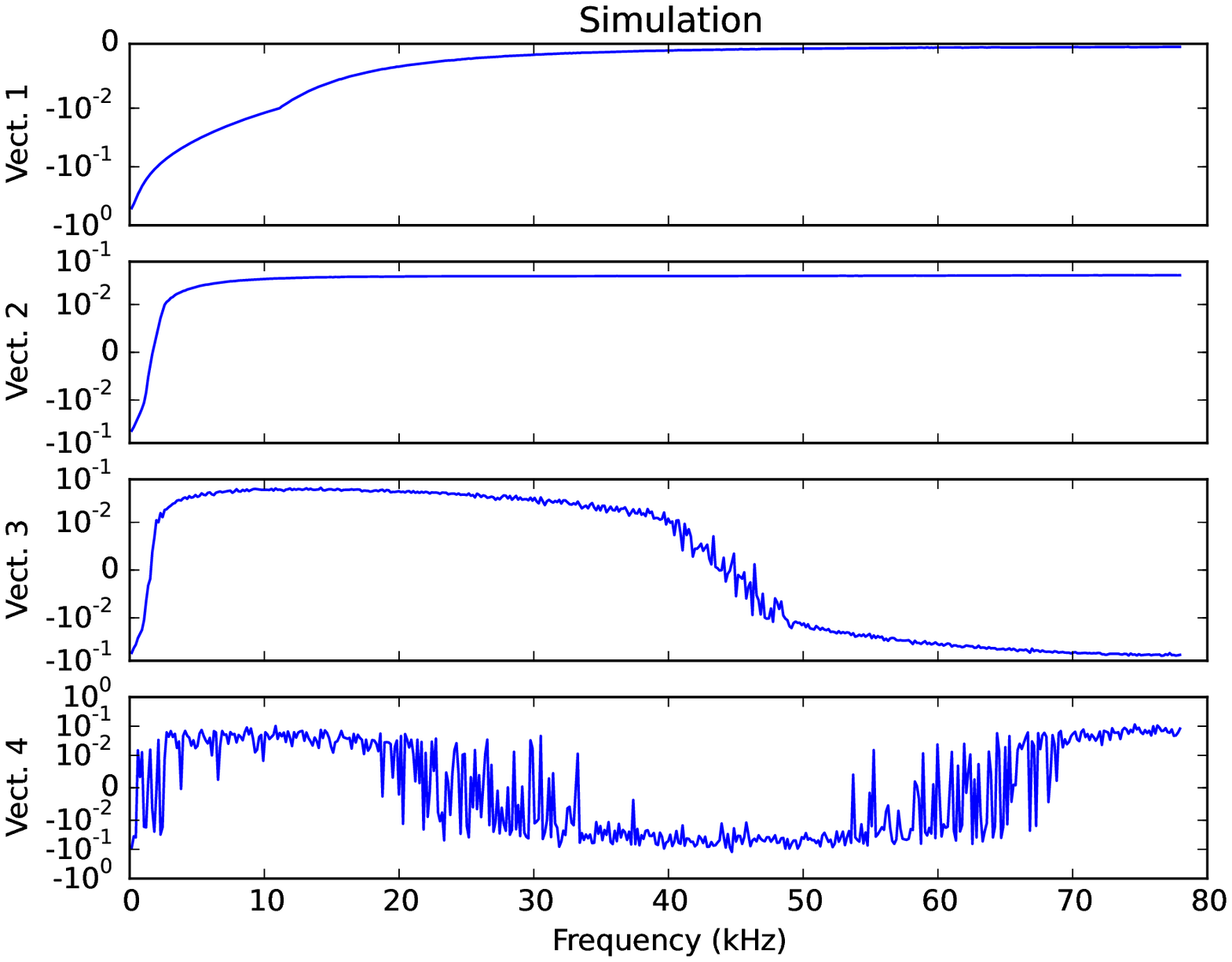}
}{
\caption{\small PCA components of pulses subject to sampling jitter, analyzed in frequency domain.}
\label{fig:f_jitt_components}
} 
\ffigbox{
\includegraphics[width=0.90\hsize,clip]{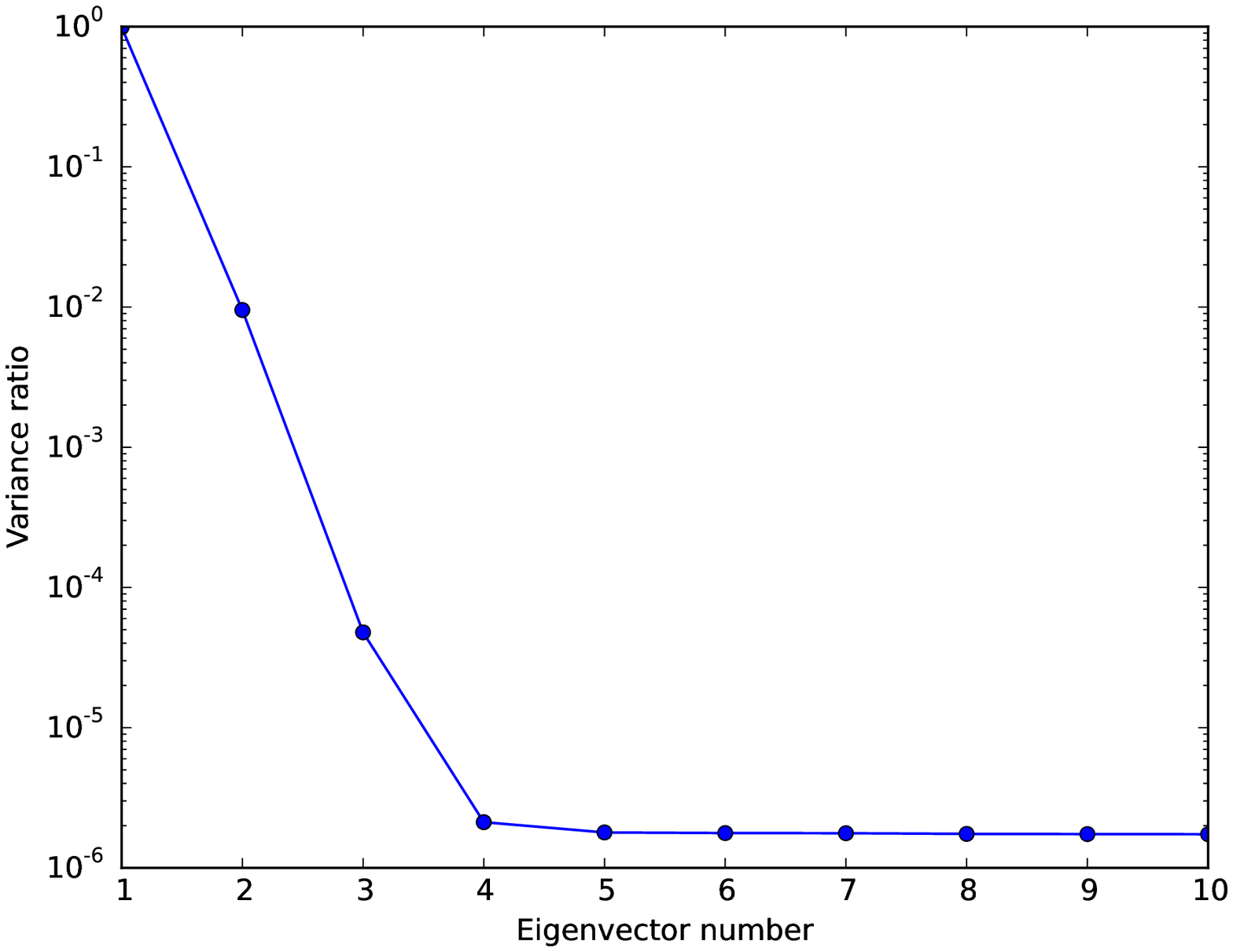}
}{
\caption{\small Variance ratio of the sampling jitter pulses in frequency domain.}
\label{fig:f_jitt_varratio}
} 
\end{floatrow}

\begin{floatrow}
\ffigbox{
\includegraphics[width=0.90\hsize,clip]{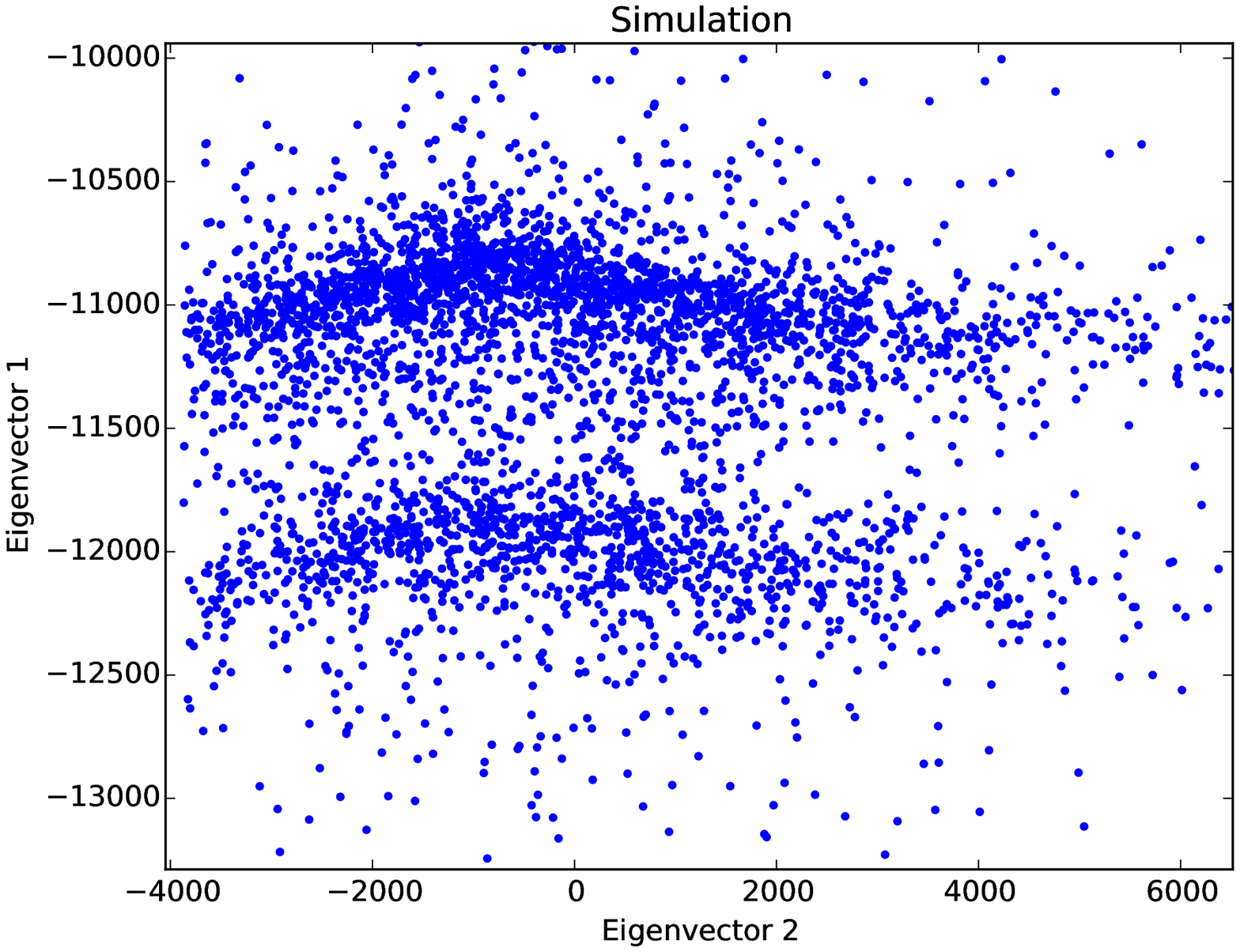}
}{
\caption{\small Correlation between the first two PCA frequency domain components.}
\label{fig:f_jitt_v1v2}
} 
\ffigbox{
\includegraphics[width=0.90\hsize,clip]{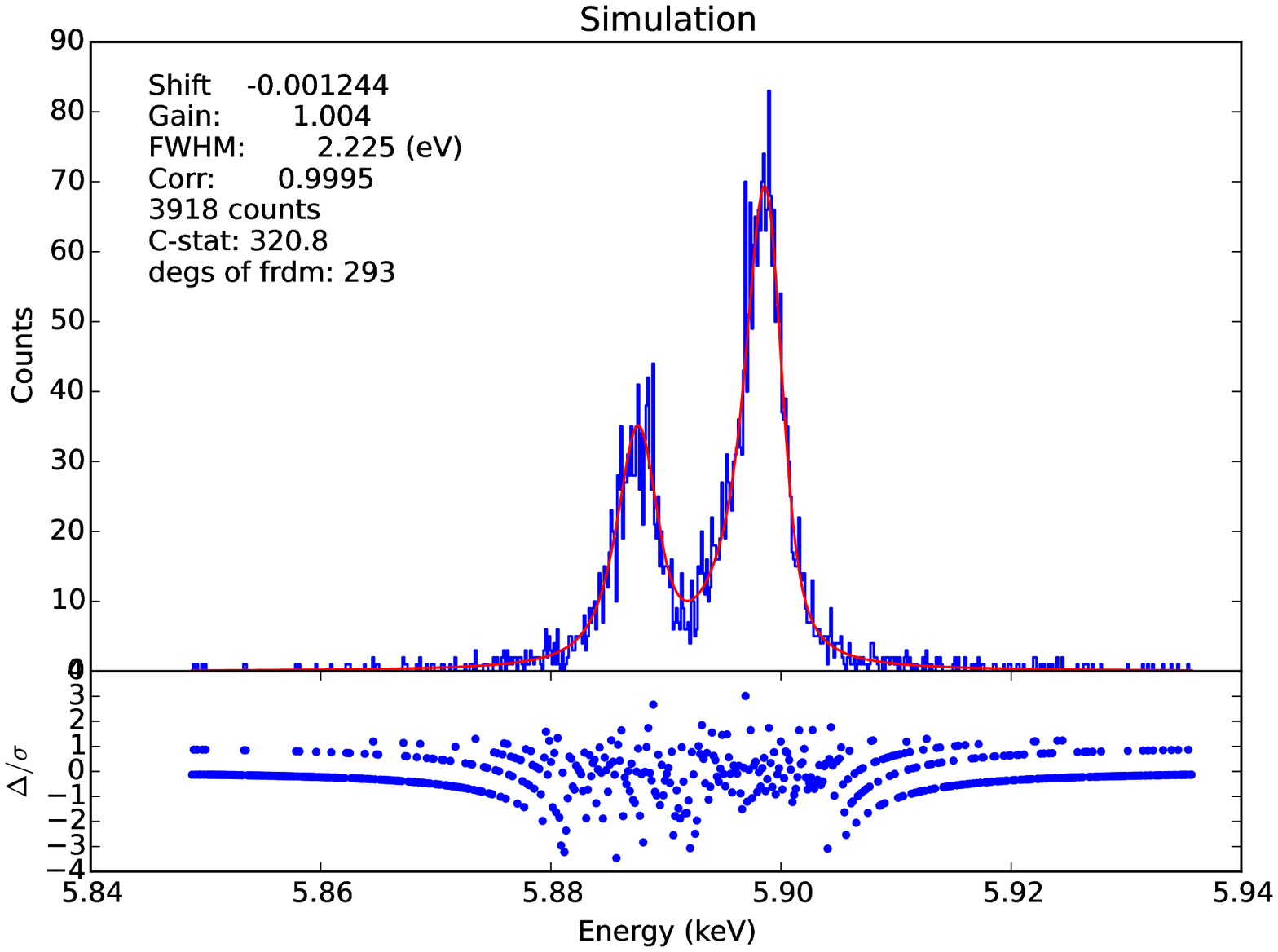}
}{
\caption{\small Fit of the instrumental resolution on the \mka line.}
\label{fig:f_jitt_lfit}
} 
\end{floatrow}
\end{figure}


The major difference is in the pulses with the sample drifts (figures \ref{fig:f_jitt_components}
to \ref{fig:f_jitt_lfit}). Since the drifts only affect the position, or phase of the pulse, in
frequency domain the effect of this drift is suppressed. The higher components are at
least a factor of 10 lower then for the time domain (fig~\ref{fig:f_jitt_varratio}).
Some correlation between the first components still exists (fig.~\ref{fig:f_jitt_v1v2}),
but the final instrumental resolution obtained is significantly better (2.2 eV, see fig.~\ref{fig:f_jitt_lfit}).
 
\section{APPLICATION TO REAL DATA}

\begin{SCfigure}
\includegraphics[width=0.45\textwidth,clip]{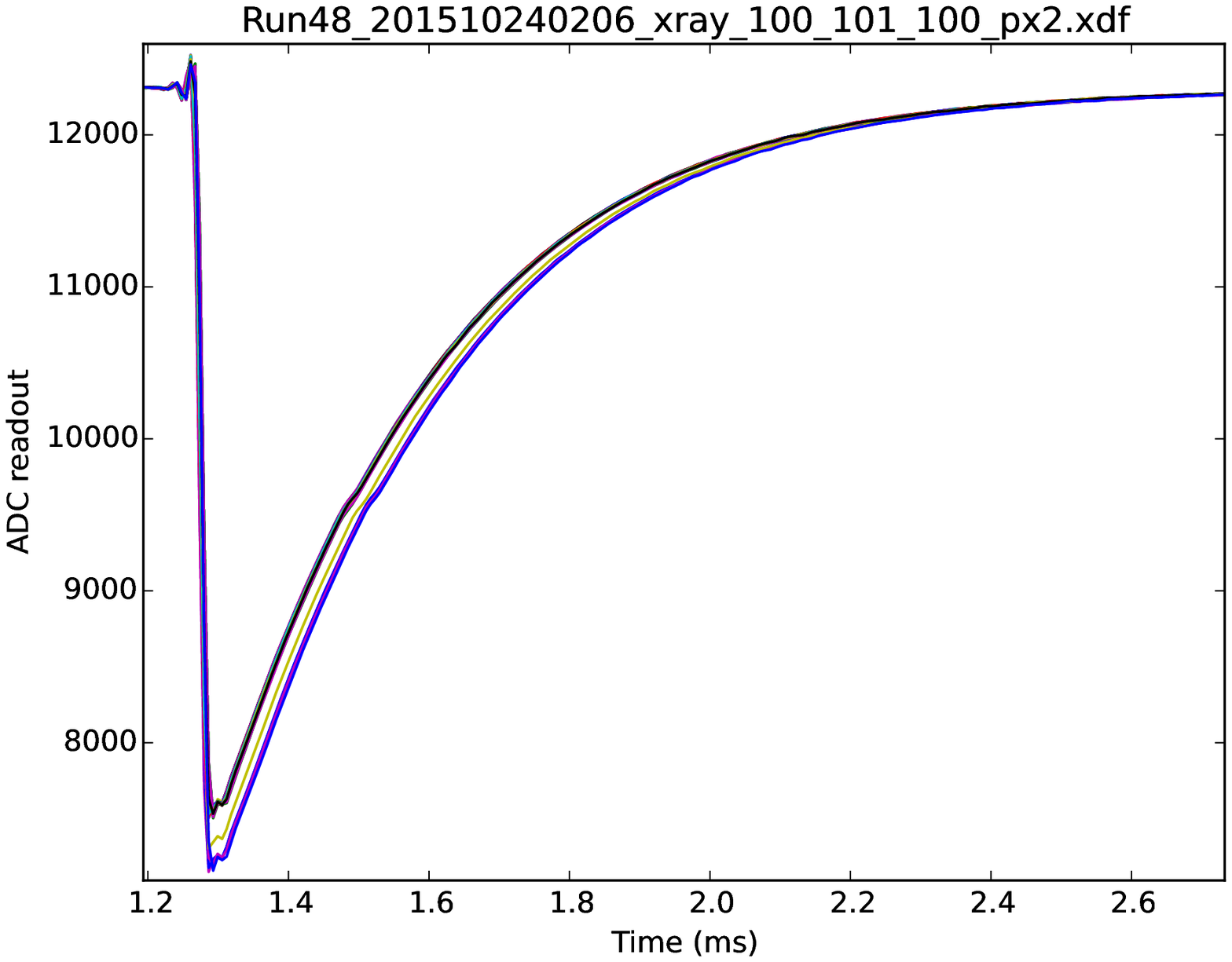}
{\caption{\small Sample of real pulses from a pixel array read out by means of frequency domain
multiplexing. The onset of the pulse is disturbed by a high frequency oscillation.  
}
\label{fig:data_sample}}
\end{SCfigure}    
 
A set of data was obtained from one pixel within a pixel array, on which multiple pixels were 
read out by means of frequency domain multiplexing. The readout circuit was not ideal, in the sense 
that the onset of the pulse was disturbed by a high frequency oscillation (see figure~\ref{fig:data_sample}).
This cause of this is still under investigation.
The noise NEP of the signal was around 2.3~eV. 

\begin{figure}[tb]
\centering
\begin{floatrow}
\ffigbox{
\includegraphics[width=0.90\hsize,clip]{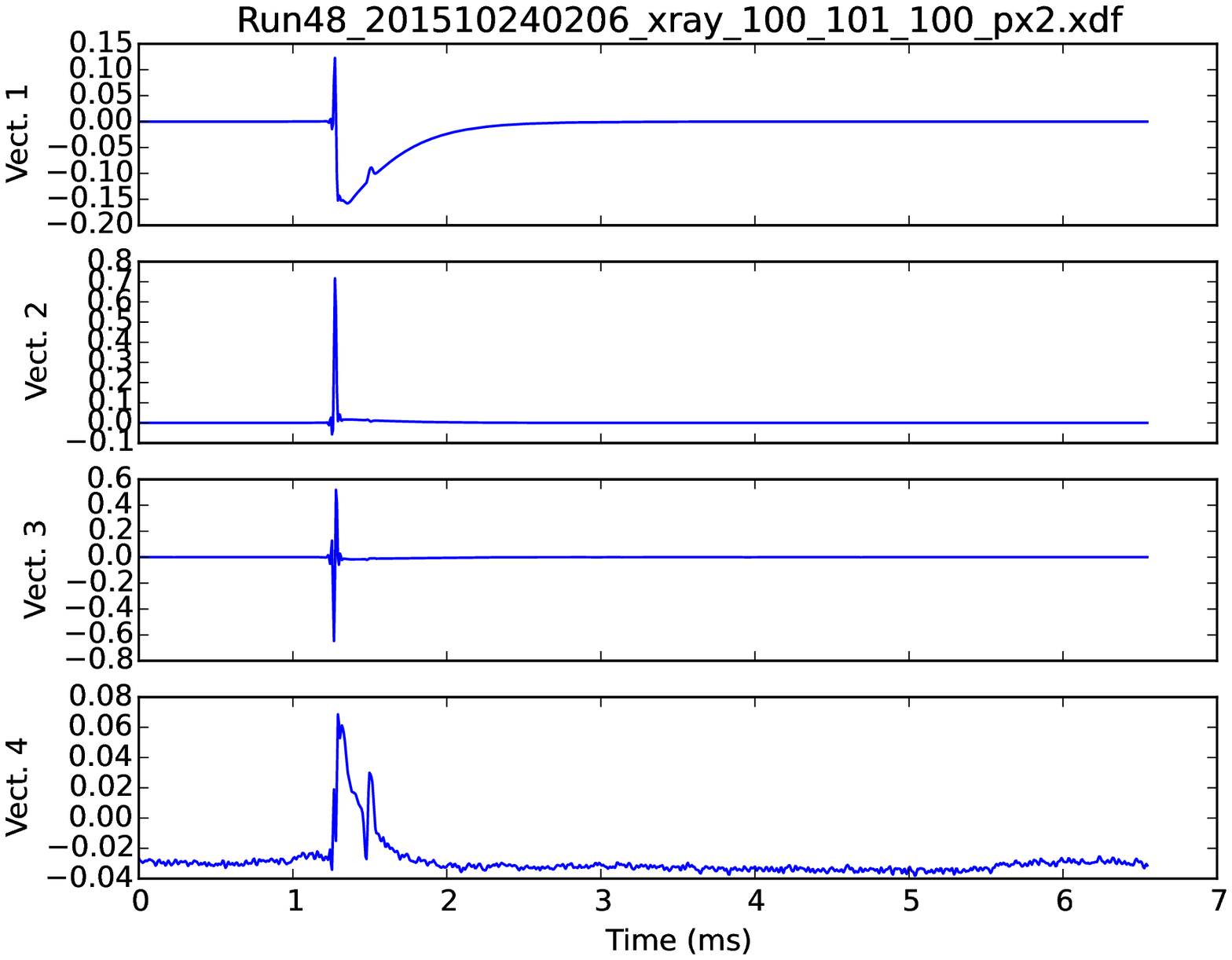}
}{
\caption{\small PCA components of real data pulses in time domain.}
\label{fig:data_components}
} 
\ffigbox{
\includegraphics[width=0.90\hsize,clip]{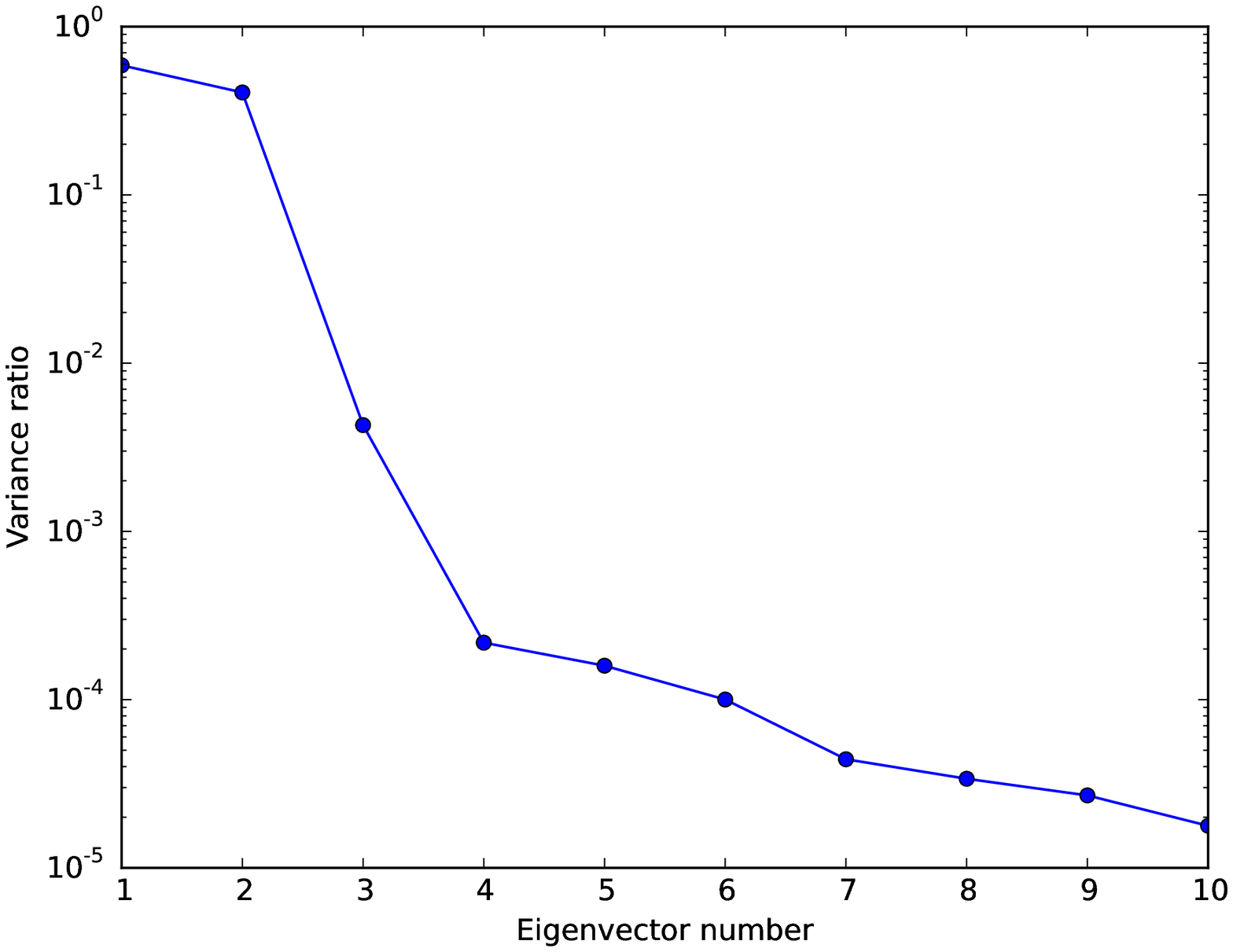}
}{
\caption{\small Variance ratio of the real data pulse components}
\label{fig:data_varratio}
} 
\end{floatrow}

\begin{floatrow}
\ffigbox{
\includegraphics[width=0.90\hsize,clip]{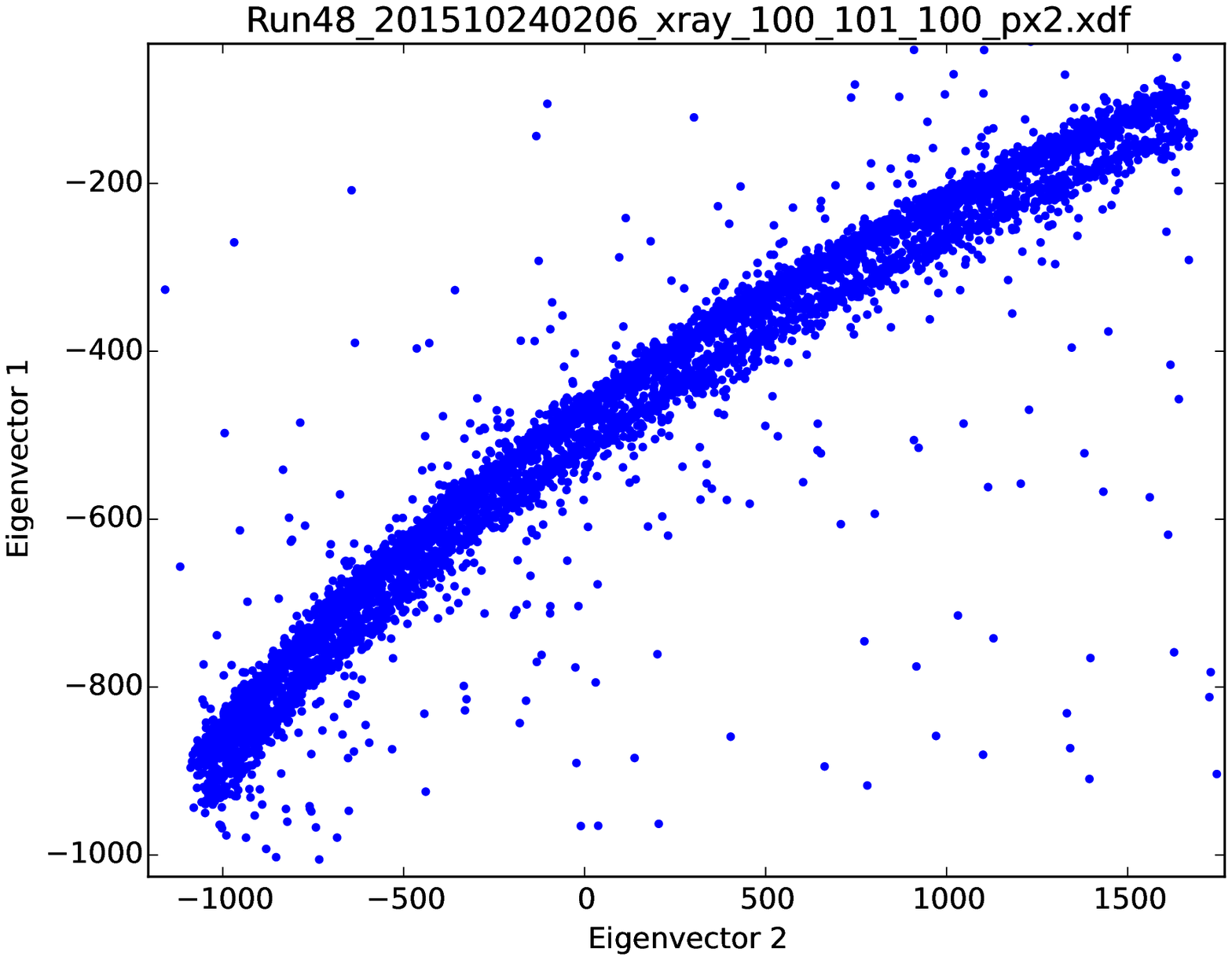}
}{
\caption{\small Correlation between the first two PCA components.}
\label{fig:data_v1v2}
} 
\ffigbox{
\includegraphics[width=0.90\hsize,clip]{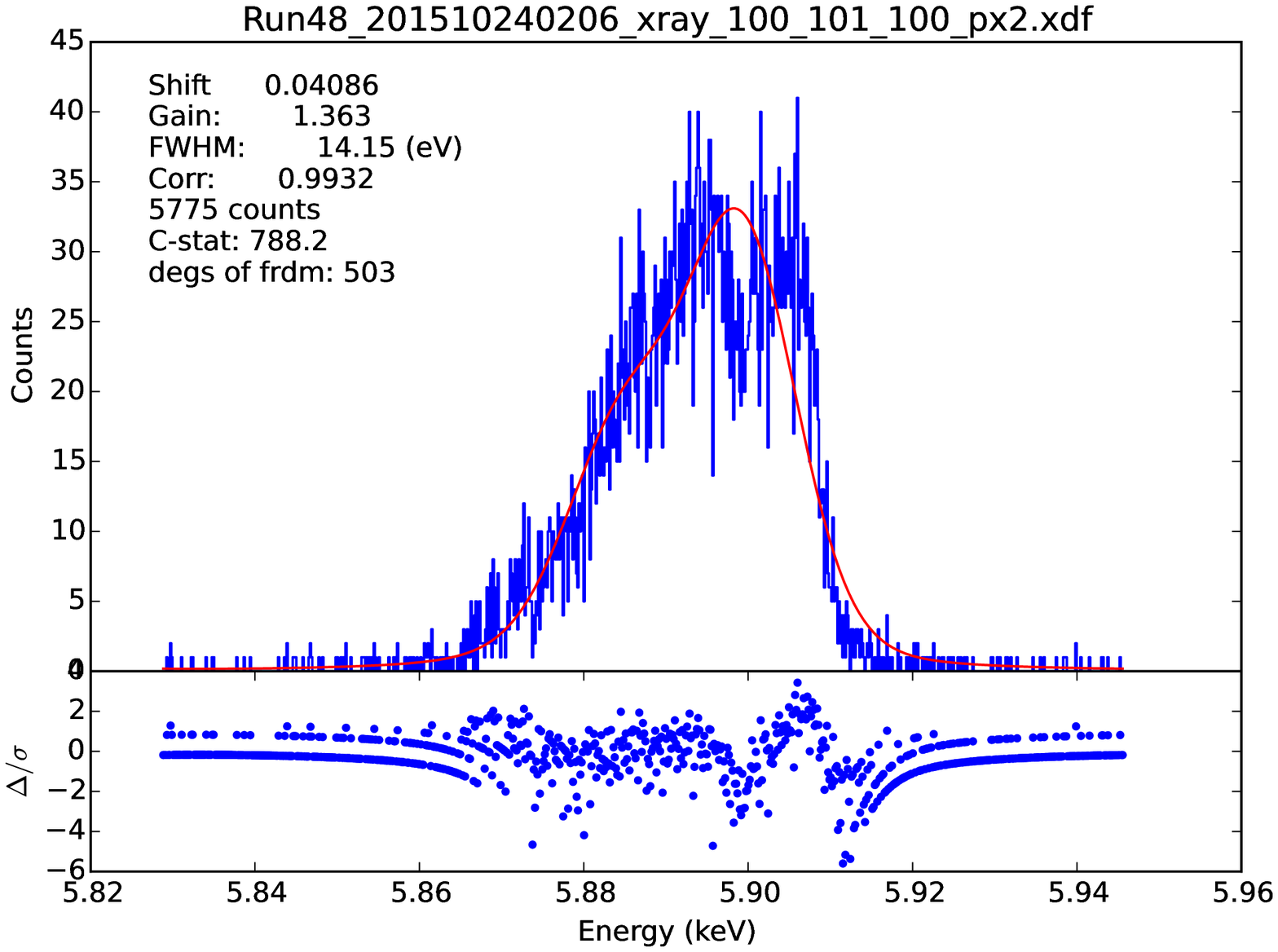}
}{
\caption{\small Fit of the instrumental resolution on the \mka line, on the time domain PCA analysis.
Clearly this time domain analysis shows bad instrumental resolution.}
\label{fig:data_lfit}
} 
\end{floatrow}

\end{figure}

\begin{figure}

\begin{floatrow}
\ffigbox{
\includegraphics[width=0.90\hsize,clip]{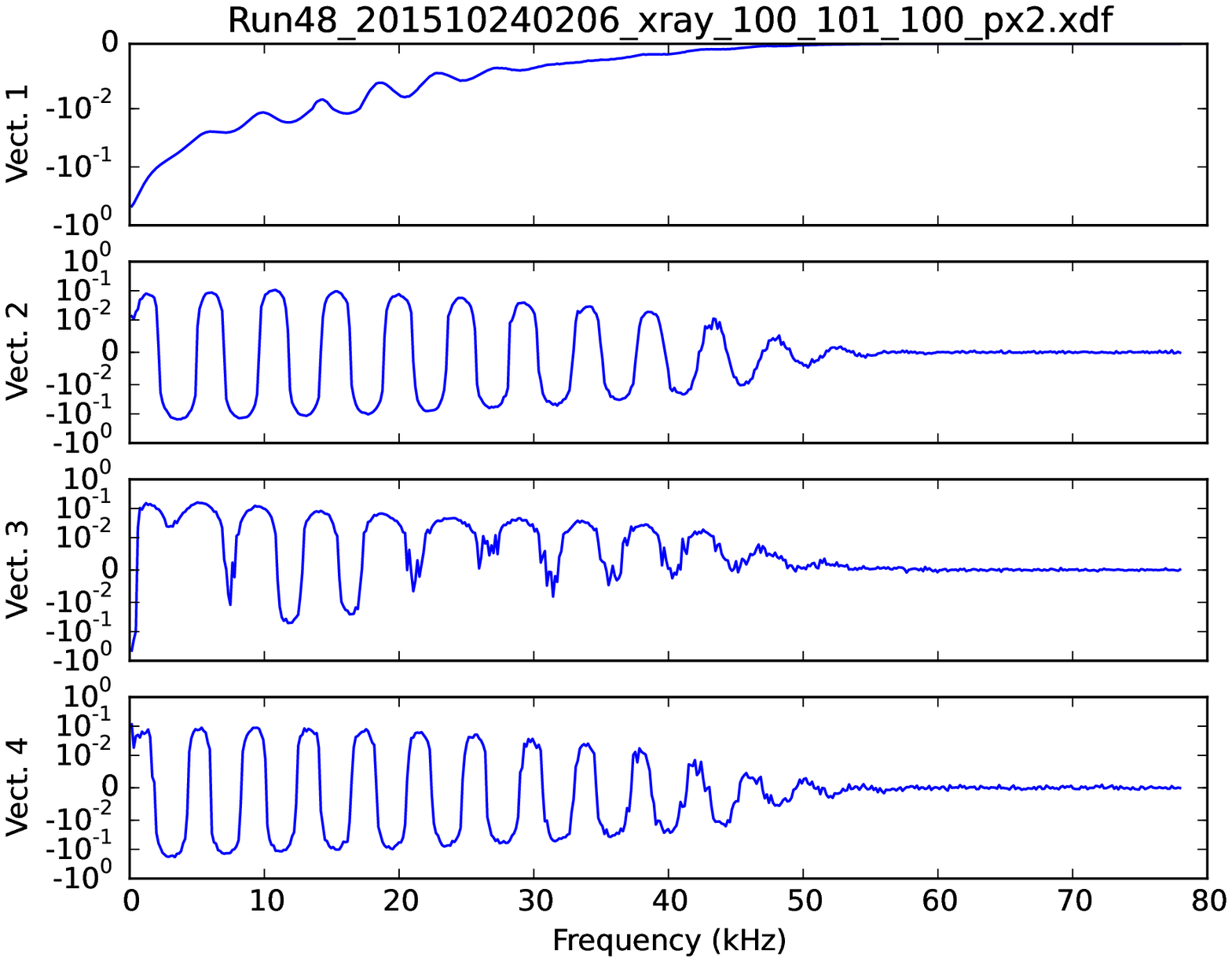}
}{
\caption{\small PCA components of real data pulses in frequency domain.}
\label{fig:f_data_components}
} 
\ffigbox{
\includegraphics[width=0.90\hsize,clip]{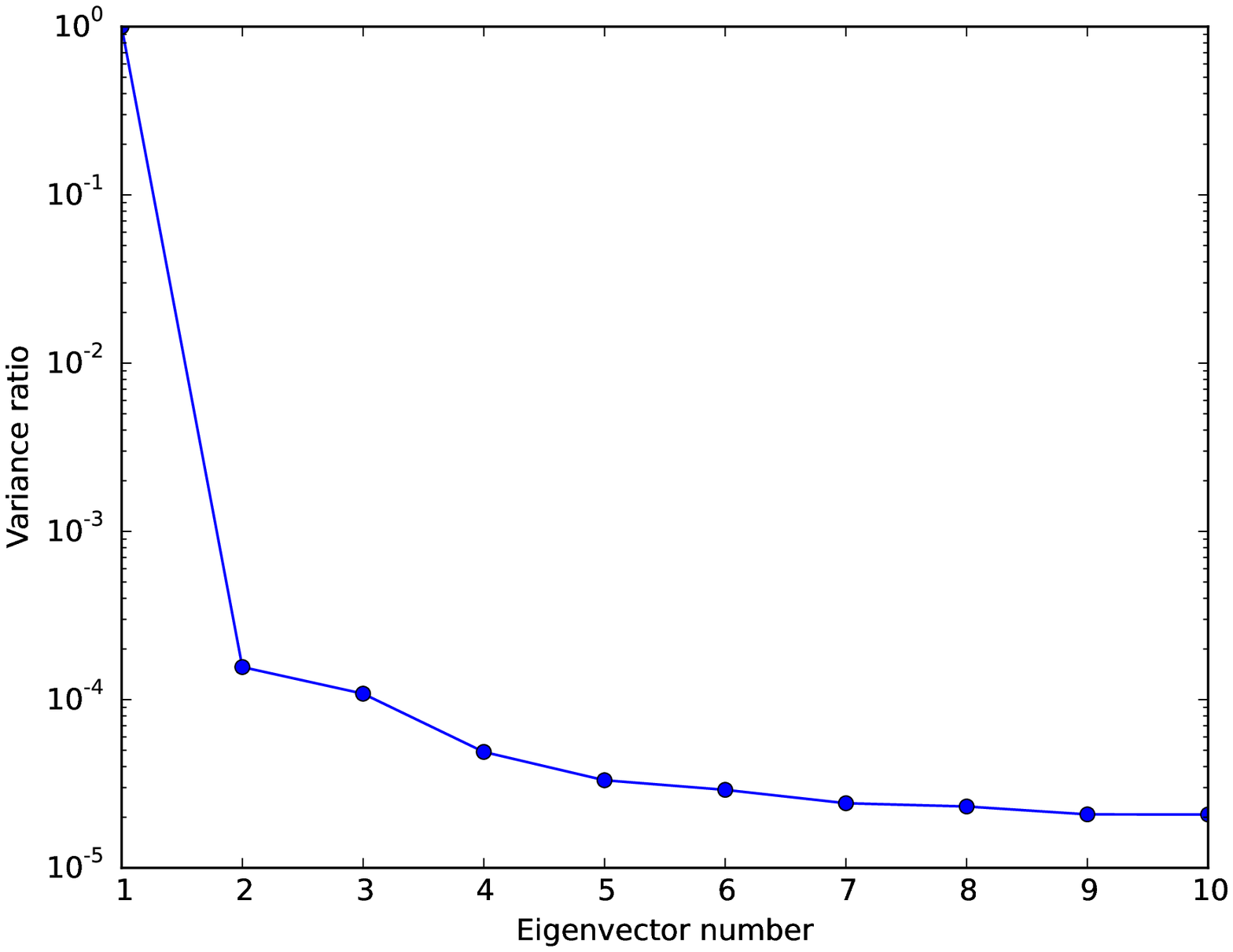}
}{
\caption{\small Variance ratio of the real data pulses in frequency domain}
\label{fig:f_data_varratio}
} 
\end{floatrow}

\begin{floatrow}
\ffigbox{
\includegraphics[width=0.90\hsize,clip]{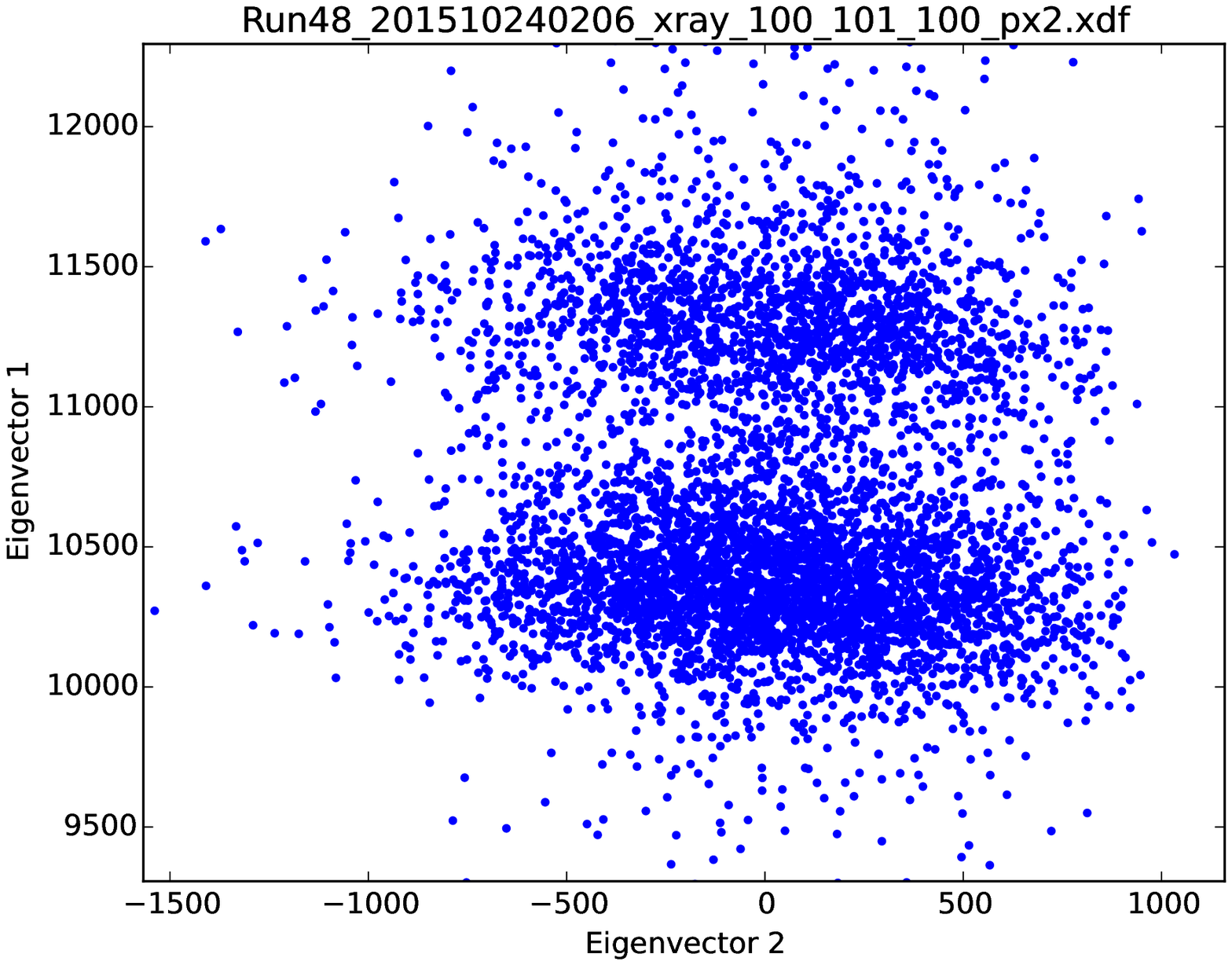}
}{
\caption{\small Correlation between the first two PCA components in frequency domain.}
\label{fig:f_data_v1v2}
} 
\ffigbox{
\includegraphics[width=0.90\hsize,clip]{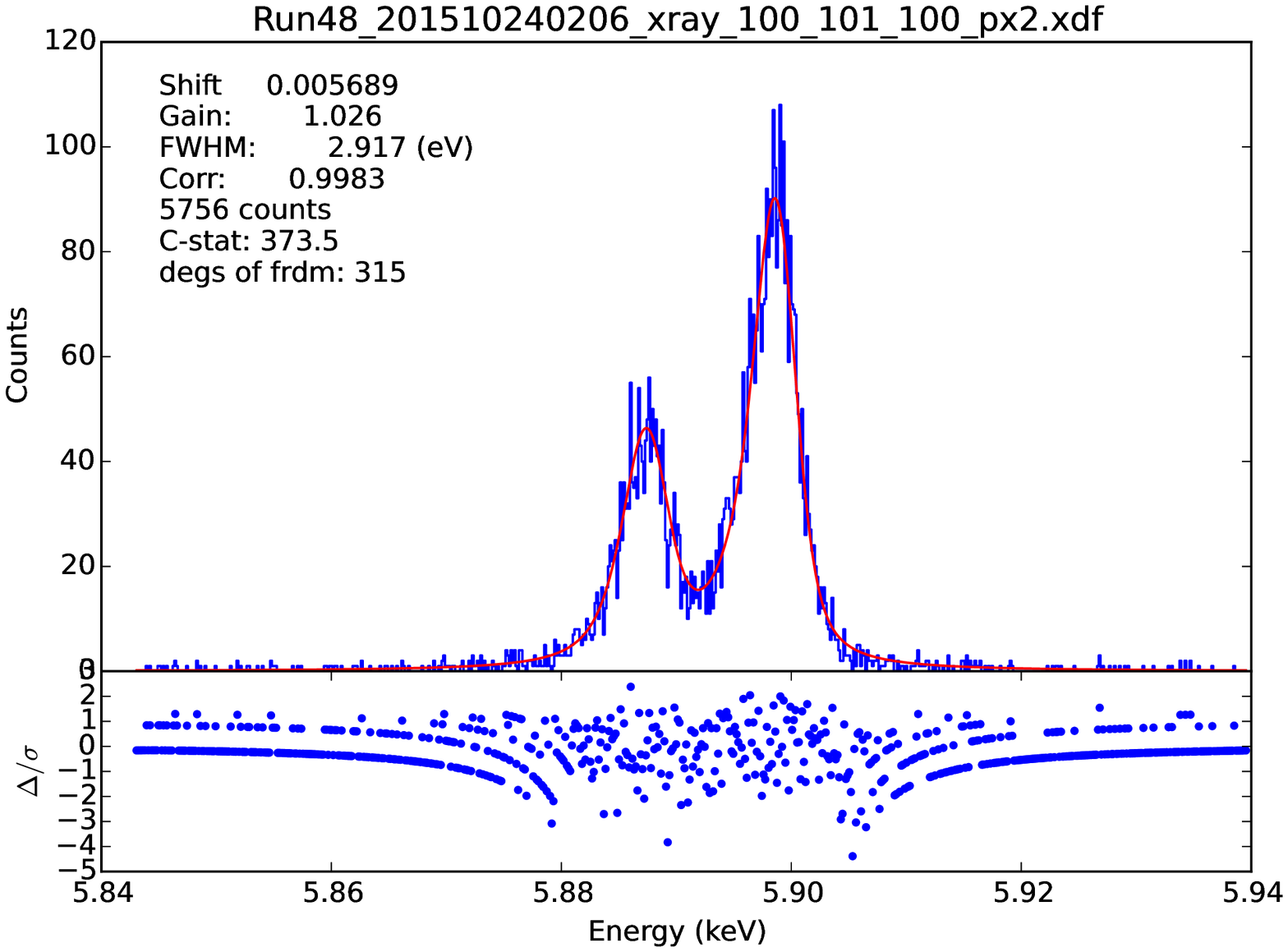}
}{
\caption{\small Fit of the instrumental resolution on the \mka line on the frequency domain PCA analysis.
The instrumental resolution obtained shows a great improvement with respect to the time domain 
analysis (fig.~\ref{fig:data_lfit}) on the same data.}
\label{fig:f_data_lfit}
} 
\end{floatrow}
\end{figure}

PCA analysis was done both in time domain and frequency domain. Figures~\ref{fig:data_components}
to \ref{fig:data_lfit} show the time domain results while figures~\ref{fig:f_data_components}
to \ref{fig:f_data_lfit} show the frequency domain results.
Here, the time domain analysis completely breaks down, while the frequency domain analysis yields
acceptable results. It appears the frequency domain PCA yields about (within the uncertainty of around 0.1 eV) the same final 
instrumental resolution as the standard optimal filtering in the frequency domain. Depending on sampling
frequency, analysis in the frequency domain does have clear advantages.



\section{CONCLUSIONS}

We performed PCA analysis on simulated data, both in time and frequency domain to study instrumental effects,
and applied this analysis on real data from a multi pixel array obtained via frequency domain multiplexing.
We conclude the following:

\begin{itemize}
\item{Different instrumental effects can be recognized in the different PCA components obtained}
\item{The relative importance of the different components in PCA analysis offers diagnostics
      on the possibility to separate different instrumental effects from photon energy}
\item{PCA clearly shows that analysis in the frequency domain directly eliminates the problems
      with sampling phase and
      offers to obtain better energy resolution in a more efficient way}
\end{itemize}


\bibliography{references}   
\bibliographystyle{spiebib}   

\end{document}